\documentclass[11pt]{article}

\usepackage[final]{acl}

\usepackage{helvet}
\usepackage{times}
\usepackage{latexsym}
\usepackage{booktabs}
\usepackage{multirow}
\usepackage{multicol}
\usepackage{amsmath}
\usepackage[table]{xcolor}
\usepackage{tabularx}
\usepackage{array}
\usepackage{algorithm}
\usepackage{algorithmic}
\usepackage{float}
\usepackage{arydshln}
\usepackage{bbding}
\usepackage{pifont}
\usepackage{subcaption}
\usepackage{colortbl}
\usepackage{makecell}
\usepackage{enumitem}
\usepackage[T1]{fontenc}
\usepackage[utf8]{inputenc}
\usepackage{microtype}
\usepackage{inconsolata}
\usepackage{graphicx}
\usepackage[most]{tcolorbox}
\usepackage{minted}

\makeatletter
\def\@afterheading{%
  \@nobreaktrue
  \everypar{%
    \if@nobreak
      \@nobreakfalse
      \clubpenalty 150\relax
      \if@afterindent \else
        {\setbox\z@\lastbox}%
      \fi
    \else
      \clubpenalty \@clubpenalty
      \everypar{}%
    \fi}}
\makeatother

\title{DSEffi-Bench: Demystifying Large Language Models' Capability in Efficient Data Science Code Generation}

\author{
Zhihao Gong$^1$\thanks{Equal contribution.},
Junzhe Yu$^2$\footnotemark[1],
Dong Huang$^4$,
Zeyu Sun$^3$,
Jie M. Zhang$^5$,
Dan Hao$^{1,2}$\thanks{Corresponding author.} \\
$^1$Key Lab of HCST (PKU), MOE; SCS, Peking University, China \\
$^2$School of Electronic and Computer Engineering, PKU Shenzhen Graduate School, China \\
$^3$Institute of Software, Chinese Academy of Sciences, Beijing, China \\
$^4$National University of Singapore, Singapore \\
$^5$King's College London, London, United Kingdom \\
zhihaogong@stu.pku.edu.cn, haodan@pku.edu.cn
}

\usepackage{albert_style}

\begin{document}
\maketitle

\begin{abstract}
Current data science (DS) code generation benchmarks equate correctness with quality, overlooking execution time differences that span orders of magnitude between correct solutions.
We introduce DSEffi-Bench, the first benchmark specifically targeting execution efficiency in LLM-generated DS code, comprising 1{,}000 instances across 10+ DS libraries with stress-testing harnesses and human-validated references.
Evaluating 16 models across 3 tiers, we find that correctness alone fails to characterize efficiency: GPT-5.4 leads in correctness (Pass, 66.9\%) but its efficiency score (B$|$P, 71.7\%) nearly matches GPT-5.4-mini (71.6\%), which solves 47 fewer tasks; Kimi-K2.5 ranks lowest in correctness among frontier models (40.2\%) yet achieves the highest efficiency score (73.6\%) across all 16 models.
A human-annotated five-category taxonomy reveals that 79.1\% of efficiency deficits extend beyond algorithmic complexity to domain-specific root causes, with distinct failure profiles across model tiers and libraries.
\revise{Two exploratory experiments provide initial evidence that these diagnostics can guide improvement, yielding up to +14.7\% efficiency gains via taxonomy-guided optimization and approaching Claude-Opus-4.6 Best@3 in efficiency at 13.0$\times$ lower cost via library-conditioned routing.}
Code and data are available at \url{https://github.com/Albert-Gong/DSEffi-Bench}.
\end{abstract}

\section{Introduction}
\label{section:introduction}

In data science (DS) coding~\cite{lai2023ds}, programmers write code to solve DS problems such as statistical modeling, numerical optimization, and deep learning.
As LLM-assisted code generation~\citep{chen2021evaluating, jain2025livecodebench} advances, DS coding has emerged as one of its most active application domains~\citep{sun2026dsaeval}.

\begin{figure}[t]
\centering
\includegraphics[width=0.9\columnwidth]{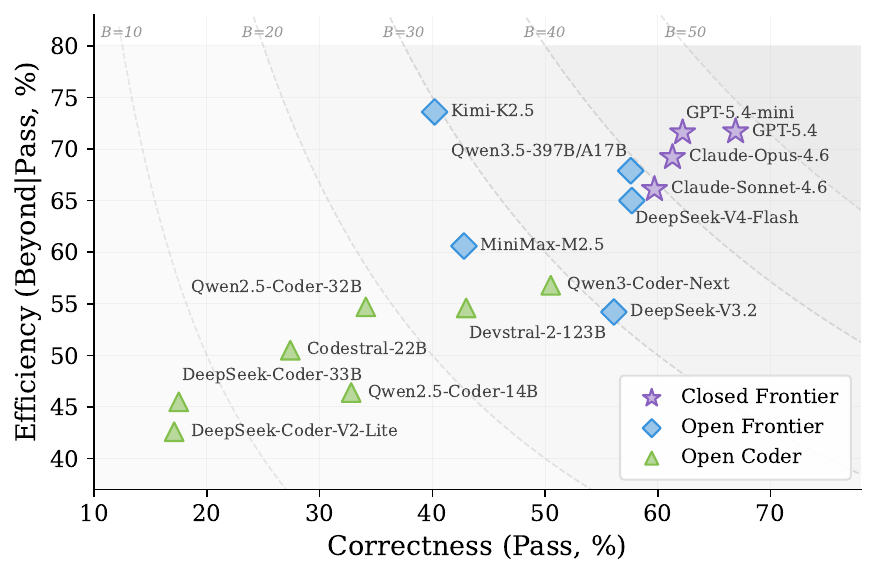}
\caption{Functional correctness (Pass, \%) vs.\ execution efficiency (B$|$P, \%) for 16~models on DSEffi-Bench.}
\label{figure:pass_bp_scatter}
\end{figure}

DS coding by nature differs from general-purpose programming.
The domain knowledge is extensive and encapsulated in highly optimized library APIs~\cite{huang2022execution,wang2026lamda}.
Writing authentic DS code demands selecting the right primitives, not merely implementing the logic from scratch.
Prior benchmarks such as ExeDS~\cite{huang2022execution}, DS-1000~\citep{lai2023ds} and DSCodeBench~\citep{ouyang2026dscodebench} curate realistic DS tasks to evaluate LLM capability.
All of them reveal uneven performance across libraries and observe that larger models tend to outperform smaller ones in correctness.

However, current benchmarks evaluate functional correctness exclusively and leave execution efficiency~\cite{niu2024evaluating, gong2026trace} unexamined.
In DS coding, efficiency depends critically on API selection: functionally equivalent paths can differ by orders of magnitude in execution time.
For instance, {\small\texttt{df.groupby('k').cummax()}} and {\small\texttt{df.groupby('k').apply(lambda g: g.cummax())}} produce identical output yet differ by an order of magnitude because the latter incurs Python-level group-wise callback overhead.
LLMs are particularly susceptible to such gaps~\citep{xue2025translibeval}, as they default to general-purpose patterns over library-specific optimized primitives. Correctness-only evaluation cannot surface this systematic weakness.

To address this gap, we introduce \textbf{DSEffi-Bench}, the first execution efficiency benchmark for evaluating LLM-based DS code generation.
DSEffi-Bench treats correctness and efficiency as two distinct capabilities and exposes their interplay through a dedicated evaluation protocol.
Concretely, we source efficiency-oriented Stack Overflow (SO) posts and apply a quality-gated four-stage pipeline that progressively filters 17{,}257 candidates down to 1{,}000 human-validated instances.
Each instance contains a problem description, an evaluation harness covering both correctness and stress test cases, and a human-validated reference solution.

Using DSEffi-Bench, we evaluate 16 models spanning 3 tiers.
First, correctness and efficiency are not aligned.
GPT-5.4 achieves the highest correctness (Pass, 66.9\%), yet its efficiency on solved tasks (B$|$P, 71.7\%) is virtually identical to GPT-5.4-mini (71.6\%), which solves 47 fewer tasks.
Conversely, Kimi-K2.5 solves the fewest tasks among frontier models (40.2\%) yet attains the highest B$|$P (73.6\%) across all 16 models.
Second, our five-category taxonomy shows that 79.1\% of efficiency failures extend beyond algorithmic complexity to domain-specific root causes.
These failures are tier-dependent: loop-to-vectorization gaps rise from 7.3\% in Closed Frontier LLMs to 29.6\% in Open Coder LLMs, while overhead-aware stack matching (OSM) declines from 29.3\% to 9.3\%.
\revise{Third, two exploratory experiments show that these diagnostics can guide improvement: taxonomy-guided optimization prompting raises efficiency by up to +14.7\%, and library-conditioned routing among open models approaches Claude-Opus-4.6 Best@3 in efficiency at 13.0$\times$ lower cost.}

This work makes the following contributions:
1)~\textbf{Benchmark.} We propose DSEffi-Bench, the first execution efficiency benchmark for DS code generation, comprising 1{,}000 human-validated instances across 10+ libraries.
2)~\textbf{Diagnosis.} We conduct a systematic evaluation and develop a five-category taxonomy that localizes 79.1\% of efficiency deficits to domain-specific root causes.
3)~\textbf{Remedy.} \revise{We explore taxonomy-guided optimization and library-aware routing and show their potential to improve efficiency.}

\section{Related Work}
\label{section:related_work}

\subsection{Efficiency Evaluation in Code Generation}

Existing efforts to evaluate the execution efficiency of LLM-generated code have largely focused on algorithmic or implementation-level optimization, leaving the DS domain underexplored.
In the competitive programming domain~\cite{puri2021codenet}, numerous benchmarks have been proposed to assess algorithmic efficiency, such as EffiBench(-X)~\citep{huang2024effibench,qing2026effibench}, Mercury~\citep{du2024mercury}, COFFE~\citep{peng2025coffe}, ENAMEL~\citep{qiu2025efficient}, and EvalPerf~\citep{liu2024evaluating}.
At the repository level~\cite{jimenez2024swe}, SWE-Perf~\citep{he2025swe} and GSO~\citep{shetty2026gso} evaluate whether models can generate patches that optimize code within the repository context.
In scientific computing, KernelBench~\citep{ouyang2025kernelbench} and AlgoTune~\citep{press2026algotune} target CUDA kernel generation and numerical algorithm redesign, respectively.

\subsection{DS Code Generation and Evaluation}

Existing DS code generation benchmarks have advanced in both evaluation harness and task scope.
ExeDS~\citep{huang2022execution} introduces execution-based evaluation for DS notebook cells, replacing surface-form metrics.
DS-1000~\citep{lai2023ds} curates 1{,}000 problems from Stack Overflow across seven libraries and combines test cases with surface-form constraints.
DSCodeBench~\citep{ouyang2026dscodebench} extends evaluation to GitHub-sourced tasks with stronger test suites across ten libraries.
DA-Code~\citep{huang2024code} and DataSciBench~\citep{zhang2025datascibench} further scale to end-to-end LLM pipeline evaluation on multi-step DS tasks.
Despite these advances, all existing work treats correctness as the sole capability proxy, leaving execution efficiency unmeasured.

\textbf{Position.}
DSEffi-Bench is the first benchmark targeting execution efficiency of LLM-generated DS code.
This dimension is orthogonal to correctness: models that scale well in correctness do not necessarily scale in efficiency, and the efficiency deficits in DS coding extend beyond algorithmic complexity to domain-specific patterns (e.g., loop-to-vectorization), distinguishing them from those studied in competitive-programming benchmarks~\citep{huang2024effibench, huang2024efficoder}.
Our evaluation confirms both points: correctness scaling~\citep{ouyang2026dscodebench} does not extend to efficiency (\S\ref{subsection:overall_evaluation}), and a human-annotated taxonomy localizes 79.1\% of deficits to library-specific root causes (\S\ref{subsection:taxonomy}).

\section{Benchmark Construction}
\label{section:benchmark_construction}

\begin{figure*}[th]
\centering
\includegraphics[width=0.9\textwidth]{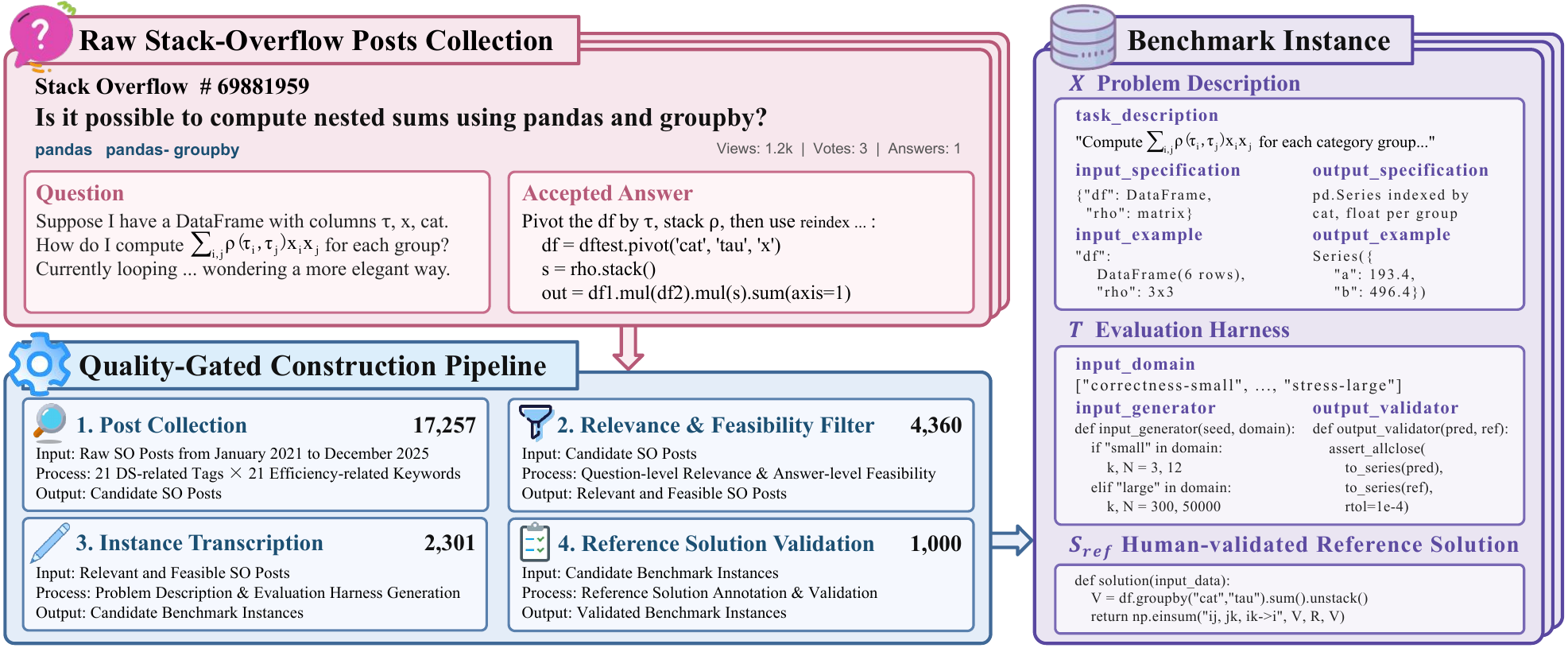}
\caption{Overview of the four-stage benchmark construction pipeline. Upper left: raw SO posts. Lower left: the construction stage. Right: the resulting benchmark instance.}
\label{figure:pipeline}
\end{figure*}

DSEffi-Bench comprises 1{,}000 instances sourced from Stack Overflow (SO), which captures everyday performance pain points that DS programmers encounter. Rather than synthesizing tasks from scratch~\citep{wu2026x,zhan2026mathsmith,wang2023self}, we derive every instance from a real SO post, preserving authentic optimization intent. To ensure that instances are both DS-related and efficiency-oriented, we design a four-stage quality-gated construction pipeline (Figure~\ref{figure:pipeline}):
(1) Efficiency-oriented Post Collection (\S\ref{subsection:efficiency_oriented_post_collection}),
(2) Relevance and Feasibility Filtering (\S\ref{subsection:relevance_and_feasibility_filtering}),
(3) Instance Transcription (\S\ref{subsection:task_transcription}), and
(4) Reference Solution Validation (\S\ref{subsection:quality_verification}).

\textbf{Benchmark Schema.}
Each instance has three components. (1) Problem description $X$, comprising five fields: task description, input specification, output specification, input example, and output example, serving as the prompt to the model under evaluation. (2) Evaluation harness $\mathcal{T}$, containing a testing domain list (\texttt{input\_domain}) that defines both correctness and stress-scale conditions, a fuzz-style input generator (\texttt{input\_generator}) for deterministic test input sampling, and a correctness checker (\texttt{output\_validator}) that compares candidate output against the reference. (3) Human-validated reference solution $S_{\text{ref}}$ to anchor both correctness judgments and efficiency scoring. The detailed schema is provided in Appendix~\ref{appendix:instance_schema}.

\subsection{Efficiency-oriented Post Collection}
\label{subsection:efficiency_oriented_post_collection}

Given the volume of SO, we first employ keyword-based retrieval via the SO Advanced Search API~\cite{stackexchange2024api} to collect posts that jointly concern a DS library and an explicit efficiency goal. We design two keyword sets targeting the two aspects: 21 DS-related library tags (e.g., Pandas, NumPy), and 21 efficiency-oriented keywords covering terms such as \textit{vectorize} and \textit{performance}. Both sets are derived from an extensive survey of prior DS benchmarks~\cite{ouyang2026dscodebench} and code efficiency studies~\cite{huang2024effibench}. Each retrieval query requires the co-occurrence of one library tag and one efficiency keyword in the same post, over January 2021 to December 2025. We retain only posts with at least 1 vote, 50 views, and an accepted answer, then deduplicate by post id, yielding 17{,}257 seed posts. The full keyword lists are in Appendix~\ref{appendix:efficiency_oriented_ds_post_collection}.

\subsection{Relevance and Feasibility Filtering}
\label{subsection:relevance_and_feasibility_filtering}

Keyword-based retrieval matches surface tokens rather than semantics~\citep{chen2024bge}: a post may mention query terms, yet neither pose a genuine efficiency problem nor offer a reproducible solution. We adopt LLM-as-Judge~\cite{li2024llms} to further assess post quality, reducing the seed set to 4{,}360 through two sequential filters: (1) question-level relevance, and (2) answer-level feasibility. 

We use Claude-Sonnet-4.6 as the judge for both filters; each returns a structured JSON decision. The question-level relevance filter takes as input the post's title and body, and asks the judge to verify that the post (i) concerns a DS coding scenario, (ii) targets efficiency improvement, and (iii) provides sufficient detail for a self-contained instance; posts that fail any criterion are discarded (17{,}257 $\to$ 5{,}850). The answer-level feasibility filter takes as input the full post (question and accepted answer), and checks whether the answer provides a concrete, reproducible optimization using DS libraries, rather than a conceptual explanation alone (5{,}850 $\to$ 4{,}360). Full prompts are provided in Appendix~\ref{appendix:relevance_and_feasibility_filtering}.

To calibrate LLM-as-Judge filters, we draw a random sample of $N{=}385$ per stage (95\% CI, $\pm$5\% margin) and have two authors independently label each instance as accept or reject. Cohen's $\kappa$ is 0.93 (question-level) and 0.71 (answer-level), indicating almost-perfect and substantial agreement, respectively~\citep{landis1977measurement}.

\subsection{Instance Transcription}
\label{subsection:task_transcription}

This stage takes the 4{,}360 filtered posts and transcribes them into candidate benchmark instances. Specifically, we (i)~generate the problem description $X$, (ii)~generate the evaluation harness $\mathcal{T}$, and (iii)~apply an LLM-driven quality scorer to discard under-specified transcriptions.

We use Claude-Sonnet-4.6 to read each post's question body and produce the problem description $X$. We explicitly formulate the unified function signature \texttt{def solution(input\_data):} in the prompt for ease of downstream model evaluation. Given $X$ and the original SO accepted answer as implementation context, we use the same model to generate the evaluation harness $\mathcal{T}$, following fuzz-style test generation~\cite{peng2025coffe, liu2024evaluating}, i.e., deterministically sampling diverse inputs from specified domains.

To filter low-quality cases, we apply an LLM-as-Judge that rates each instance on four dimensions (1--3 scale; 1=low quality, 2=adequate, 3=high quality): (i)~task description clarity, (ii)~I/O specification precision, (iii)~input generator correctness, and (iv)~output validator robustness. Instances where any dimension receives a score of 1 are discarded. To mitigate model self-preference bias~\cite{wataoka2024self}, we select a judge from a different model family (GPT-5.4). This yields 2{,}301 candidate instances. All transcription prompts are provided in Appendix~\ref{appendix:task_transcription}.

\subsection{Human-Validated Reference Solution}
\label{subsection:quality_verification}
This stage augments the LLM-generated artifacts with human-validated reference solutions. To balance human effort with benchmark quality, we rank candidates by their mean quality score from the previous judging process and annotate from the top down until 1{,}000 instances are obtained.

Two authors, each with over five years of DS coding experience, serve as annotators. Both annotators independently study the problem description, the I/O specification and example, and the original SO accepted answer. Each then writes a reference solution conforming to the \texttt{def solution(input\_data):} signature. Annotators may choose to implement the accepted answer's optimization suggestion faithfully or further improve it when a more efficient approach is apparent.

During annotation, 50.3\% of reference solutions adopt a fundamentally different strategy, 32.2\% further refine the answer's approach, and the remaining 17.5\% faithfully reproduce it. Afterwards, the two solutions are compared; when they differ, annotators discuss and select the more efficient one, or compose a hybrid that combines the best elements. Annotators also revise the problem description, input generator, and output validator when any artifact was ambiguous or incomplete, ensuring that the final instances are quality-gated and not simple reproductions of any single LLM's output.

In total, we annotated 1{,}247 candidates and obtained 1{,}000 validated instances; the remaining 247 were discarded due to quality reasons such as ambiguous problem statements, or tasks where the reference solution could not establish a clear efficiency advantage over naive approaches. Detailed annotation statistics are in Appendix~\ref{appendix:annotation_statistics}.

\textbf{Benchmark Statistics.} As summarized in Table~\ref{table:benchmark_glance}, the benchmark spans five years of SO posts (2021--2025) and covers over 10 DS libraries, including representative libraries such as NumPy, Pandas, PyTorch, Polars, and SciPy, along with Numba, TensorFlow, and other long-tail libraries. This distribution reflects the natural frequency of efficiency-oriented questions on SO, where NumPy and Pandas dominate everyday DS workloads.

\begin{table}[t]
\centering
\footnotesize
\caption{Benchmark statistics and evaluation protocol.}
\label{table:benchmark_glance}
\begin{tabular}{@{}l r@{}}
\toprule
\textbf{Statistic} & \textbf{Value} \\
\midrule
\multicolumn{2}{@{}>{\columncolor{gray!10}[0pt][0pt]}l@{}}{\textit{Benchmark Profile}} \\
Instances & 1{,}000 \\
SO time span & 2021.01--2025.12 \\
Problem description (words) & 358.3 \\
DS libraries & 10+ \\
\midrule
\multicolumn{2}{@{}>{\columncolor{gray!10}[0pt][0pt]}l@{}}{\textit{Evaluation Protocol}} \\
Evaluated models & 16 (3 tiers) \\
Reference solution (lines) & 27.5 \\
Test cases (per instance) & 38.8 \\
Spectrum size (mean / median) & 29.4 / 31.0 \\
Slowest/fastest ratio (median) & 14.5$\times$ \\
\bottomrule
\end{tabular}
\end{table}

\section{Evaluation Protocol}
\label{section:evaluation_protocol}

Given a problem description~$X$, a model generates a candidate solution~$S_c$. We evaluate each~$S_c$ along two dimensions using the evaluation harness~$\mathcal{T}$ and the human-validated reference~$S_{\text{ref}}$: functional correctness and execution efficiency. Table~\ref{table:benchmark_glance} summarizes the evaluation protocol.

\subsection{Evaluation Metrics}
\label{subsection:beyond_metric}

\textbf{Functional correctness (Pass).}
For each task, we use \texttt{input\_generator} to produce multiple test inputs $\{x_i\}$. $S_c$ is correct if and only if $V\!\bigl(S_c(x_i),\, S_{\text{ref}}(x_i)\bigr) = \textsc{True}$ for every~$x_i$, where $V$ is the \texttt{output\_validator}. Solutions that fail or exceed the 5-second timeout on any test case receive a score of 0. \textbf{Pass} is the fraction of tasks where the model passed all the tests.

\textbf{Execution efficiency (Beyond, Beyond|Pass).}
Following Mercury~\citep{du2024mercury}, we evaluate efficiency via a relative score. For each task, we construct an efficiency spectrum from a pool of correct solutions sorted by execution time. Each solution receives a normalized score $\text{Beyond}(S_c) = (\log t_{\max} - \log t_c)\,/\,(\log t_{\max} - \log t_{\min})$, where $t_c$ is the execution time of $S_c$, and $t_{\min}$, $t_{\max}$ are the fastest and slowest in the spectrum. The fastest solution scores 1; slowest and incorrect solutions receive 0.
Note that the original Beyond metric in Mercury uses linear normalization. We find that execution time spans in our benchmark are wide (median slowest-to-fastest ratio 14.5$\times$, 90th percentile 522.0$\times$), causing 46.6\% of correct solutions to score above 0.95 under linear normalization, which compresses meaningful differences into a narrow band. We therefore normalize in log-space, which spreads the median score to 0.74 and preserves discriminative range. 

We report three model-level metrics: \textbf{Pass} (fraction of tasks solved), \textbf{Beyond} (mean spectrum score, 0 for unsolved), and \textbf{Beyond|Pass (B$|$P)} (mean spectrum score restricted to solved tasks, isolating efficiency from correctness).

\subsection{Efficiency Spectrum}
\label{subsection:models_and_generation}

We evaluate 16~models grouped into 3 tiers by accessibility and specialization: \textbf{Closed Frontier} (proprietary general-purpose, e.g., GPT-5.4, Claude-Opus-4.6), \textbf{Open Frontier} (open-weight general-purpose, e.g., DeepSeek-V4-Flash, Kimi-K2.5), and \textbf{Open Coder} (open-weight code-specialized, e.g., Qwen3-Coder-Next, Devstral-2-123B). The full model list is in Appendix~\ref{section:appendix_a3}.

The efficiency spectrum is constructed from all correct solutions produced by these 16~models plus the human reference. Each model generates one greedy sample (T{=}0.0) and three temperature samples (T{=}0.8, following~\citet{liu2024evaluating}) per task, contributing up to four correct solutions per model per task. This yields an average spectrum size of 29.4 per task (median 31.0), with a median slowest-to-fastest ratio of 14.5$\times$ (90th percentile: 522.0$\times$), providing sufficient dynamic range for efficiency discrimination.

\subsection{Measurement Procedure}
\label{subsection:correctness_and_efficiency_measurement}
The \texttt{input\_generator} (described in \S\ref{subsection:task_transcription}) produces on average 38.8~test cases per task (26.8~correctness, 12.0~stress). For correct solutions, we measure execution time via Python's \texttt{time.perf\_counter} using a repeated-execution protocol: 1 warm-up run followed by 3 timed runs, reporting the mean elapsed time. A solution's execution time for a task is the average across tests. The full prompts and the rationale for the measurement parameters are in Appendix~\ref{section:appendix_a3}.

\section{Evaluation}
\label{section:evaluation}

\begin{table*}[t]
\centering
\caption{Model performance on DSEffi-Bench (T{=}0.0). \textbf{Pass (abbr. P)}: tasks solved. \textbf{Beyond (abbr. B)}: normalized efficiency score. \textbf{B$|$P}: Beyond restricted to passed tasks. \textbf{Bold} = best in tier; \underline{underline} = second.}
\label{table:main_results}
\renewcommand{\arraystretch}{1.00}
\resizebox{1.0\textwidth}{!}{%
\scriptsize
\setlength{\tabcolsep}{3pt}
\definecolor{tblue}{RGB}{232,244,255}
\newcolumntype{B}{>{\columncolor{tblue}}r}
\begin{tabular}{l >{\columncolor{tblue}}w{r}{3.6em} >{\columncolor{tblue}}w{r}{3.6em} >{\columncolor{tblue}}w{r}{3.6em} @{\hspace{8pt}{\color{gray!40}\vrule width 0.5pt}\hspace{8pt}} rr BB rr BB rr BB}
\toprule
 & \multicolumn{3}{c}{\textbf{Overall}} & \multicolumn{2}{c}{\textbf{NumPy}} & \multicolumn{2}{c}{\textbf{Pandas}} & \multicolumn{2}{c}{\textbf{PyTorch}} & \multicolumn{2}{c}{\textbf{SciPy}} & \multicolumn{2}{c}{\textbf{Polars}} & \multicolumn{2}{c}{\textbf{Others}} \\
\cmidrule(lr){2-4} \cmidrule(lr){5-6} \cmidrule(lr){7-8} \cmidrule(lr){9-10} \cmidrule(lr){11-12} \cmidrule(lr){13-14} \cmidrule(lr){15-16}
\rowcolor{white}
\textbf{Model} & \textbf{Pass} & \textbf{Beyond} & \textbf{B$|$P} & \textbf{P} & \textbf{B$|$P} & \textbf{P} & \textbf{B$|$P} & \textbf{P} & \textbf{B$|$P} & \textbf{P} & \textbf{B$|$P} & \textbf{P} & \textbf{B$|$P} & \textbf{P} & \textbf{B$|$P} \\
\midrule
\multicolumn{16}{@{}>{\columncolor{gray!10}[0pt][0pt]}l@{}}{\textit{Closed Frontier}} \\
\cmidrule[\lightrulewidth]{1-16}
GPT-5.4 & \textbf{66.9} & \textbf{48.0} & \textbf{71.7} & \textbf{76.8} & \textbf{78.9} & \textbf{65.3} & \underline{71.2} & 64.7 & \underline{53.6} & 64.6 & \textbf{82.3} & \textbf{72.6} & \underline{59.7} & \underline{37.6} & 48.6 \\
GPT-5.4-mini & \underline{62.2} & \underline{44.6} & \underline{71.6} & \underline{71.4} & \underline{77.2} & 58.0 & \textbf{71.9} & \textbf{73.5} & \textbf{60.9} & \underline{70.8} & 67.3 & 62.9 & 57.4 & 32.5 & \textbf{62.0} \\
Claude-Opus-4.6 & 61.3 & 42.4 & 69.2 & 64.9 & 73.4 & \underline{59.6} & 69.6 & \underline{67.6} & 49.4 & \textbf{75.0} & \underline{80.3} & \textbf{72.6} & \textbf{66.9} & \textbf{38.5} & \underline{57.1} \\
Claude-Sonnet-4.6 & 59.7 & 39.4 & 66.1 & 67.3 & 68.3 & 57.7 & 69.6 & 55.9 & 48.2 & 62.5 & 74.6 & 67.7 & 57.4 & 36.8 & 55.8 \\
\midrule
\multicolumn{16}{@{}>{\columncolor{gray!10}[0pt][0pt]}l@{}}{\textit{Open Frontier}} \\
\cmidrule[\lightrulewidth]{1-16}
DeepSeek-V4-Flash & \textbf{57.7} & \underline{37.5} & 65.0 & 65.5 & \underline{70.7} & \underline{54.9} & 63.2 & \textbf{58.8} & 40.4 & \textbf{68.8} & 67.6 & 46.8 & \textbf{72.1} & \textbf{40.2} & 55.5 \\
DeepSeek-V3.2 & 56.1 & 30.4 & 54.2 & \underline{66.0} & 59.9 & 52.7 & 52.3 & \underline{57.4} & 46.2 & \underline{58.3} & 64.2 & 51.6 & 41.1 & 33.3 & 36.8 \\
Kimi-K2.5 & 40.2 & 29.6 & \textbf{73.6} & 44.6 & \textbf{79.6} & 41.0 & \textbf{74.0} & 41.2 & \textbf{59.1} & 27.1 & \textbf{87.6} & \underline{53.2} & 58.1 & 21.4 & \textbf{60.0} \\
MiniMax-M2.5 & 42.8 & 26.0 & 60.6 & 47.9 & 64.4 & 43.5 & 60.3 & 30.9 & 35.5 & 47.9 & 64.9 & 40.3 & 56.0 & 29.9 & \underline{57.4} \\
Qwen3.5-397B/A17B & \underline{57.6} & \textbf{39.1} & \underline{67.9} & \textbf{67.0} & 70.2 & \textbf{57.1} & \underline{72.2} & 52.9 & \underline{51.0} & 43.8 & \underline{69.8} & \textbf{58.1} & \underline{62.0} & \underline{35.9} & 53.3 \\
\midrule
\multicolumn{16}{@{}>{\columncolor{gray!10}[0pt][0pt]}l@{}}{\textit{Open Coder}} \\
\cmidrule[\lightrulewidth]{1-16}
Qwen3-Coder-Next & \textbf{50.5} & \textbf{28.7} & \textbf{56.8} & \textbf{59.5} & \textbf{58.9} & \textbf{46.7} & \textbf{57.1} & \textbf{55.9} & 53.2 & \textbf{52.1} & 54.5 & \textbf{38.7} & 56.8 & \textbf{33.3} & \underline{48.4} \\
Devstral-2-123B & \underline{43.0} & \underline{23.5} & 54.6 & \underline{52.3} & \underline{57.8} & \underline{39.7} & \underline{55.8} & 35.3 & 40.2 & 37.5 & \textbf{64.1} & \underline{35.5} & 51.0 & \underline{31.6} & 40.3 \\
Qwen2.5-Coder-32B & 34.1 & 18.7 & \underline{54.7} & 40.7 & 56.5 & 31.5 & 52.2 & 35.3 & \underline{73.3} & 41.7 & 47.4 & 14.5 & \textbf{69.9} & 25.6 & 39.3 \\
Qwen2.5-Coder-14B & 32.8 & 15.2 & 46.4 & 39.9 & 44.0 & 29.7 & 50.6 & 33.8 & 53.2 & \underline{43.8} & 39.9 & 11.3 & \underline{61.0} & 23.9 & 41.9 \\
Codestral-22B & 27.4 & 13.8 & 50.5 & 34.5 & 49.5 & 23.0 & 53.5 & \underline{39.7} & 44.0 & 29.2 & 61.1 & 12.9 & 55.2 & 15.4 & 44.4 \\
DeepSeek-Coder-33B & 17.5 & 8.0 & 45.5 & 19.3 & 47.2 & 16.1 & 45.2 & 26.5 & 42.4 & 22.9 & \underline{61.7} & 11.3 & 39.5 & 11.1 & 30.9 \\
DeepSeek-Coder-V2-Lite & 17.1 & 7.3 & 42.6 & 20.1 & 32.5 & 12.3 & 36.3 & 27.9 & \textbf{84.2} & 20.8 & 31.1 & 12.9 & 50.5 & 14.5 & \textbf{60.3} \\
\midrule
\textbf{Human Reference} & 100.0 & 86.6 & 86.6 & 100.0 & 88.5 & 100.0 & 85.8 & 100.0 & 78.7 & 100.0 & 94.2 & 100.0 & 78.1 & 100.0 & 88.3 \\
\bottomrule
\end{tabular}%
}
\end{table*}

This section presents evaluation results of 16~LLMs along correctness and efficiency.

\subsection{Overall Evaluation}
\label{subsection:overall_evaluation}


Table~\ref{table:main_results} and Figure~\ref{figure:pass_bp_scatter} summarize the main results. \textbf{Correctness alone fails to characterize efficiency.} Table~\ref{table:main_results} shows a partial decoupling between correctness and efficiency.
GPT-5.4 leads in correctness (Pass, 66.9\%), yet its efficiency score (B$|$P, 71.7\%) is nearly tied with GPT-5.4-mini (71.6\%), which solves 47 fewer tasks.
The gap widens for open models. Kimi-K2.5 ranks lowest (40.2\%) in Pass among the frontier models, yet achieves the highest B$|$P (73.6\%) across all models.
DeepSeek-V3.2 presents the mirror image, with Pass 56.1\% but B$|$P only 54.2\%, trailing Kimi by 19.4\%. These rank inversions contrast with the scaling observation reported by correctness-only DS benchmarks~\citep{lai2023ds, ouyang2026dscodebench}.

Moreover, code-specialized training does not close this gap. Open Coder models average B$|$P 50.2\%, trailing Open Frontier generalists (64.3\%) despite explicit code-corpus training.
Meanwhile, the human reference achieves B$|$P = 86.6\%, outperforming the best LLM by 13.0\%, suggesting that generating efficient DS code remains far from saturated for current LLMs.

\begin{figure}[t]
\centering
\includegraphics[width=.95\linewidth]{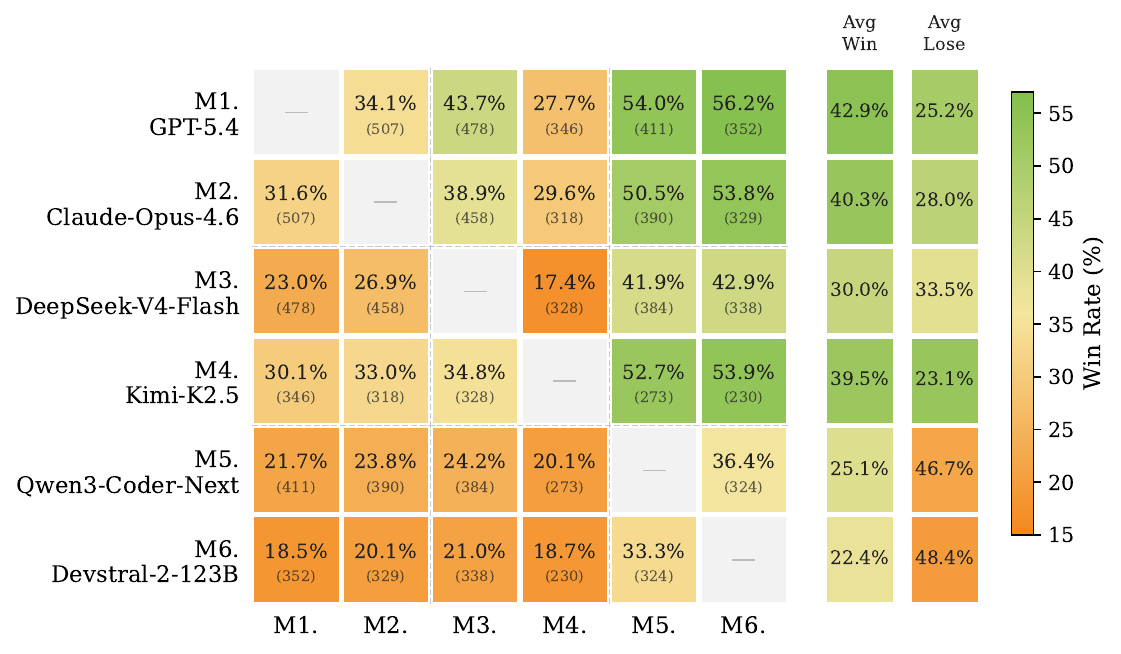}
\caption{Pairwise efficiency comparison of six representative models.}
\label{figure:hero_pairwise}
\end{figure}

\textbf{Head-to-head comparison sharpens the picture.}
To move beyond aggregate metrics, we compute pairwise efficiency win rates on models' common-correct subsets.
For each model pair $(A, B)$, we restrict to tasks that both solve and define $A$ as more efficient (winning) if $(t_B - t_A)/t_B \geq 0.10$ following~\citet{shypula2023learning}.
Figure~\ref{figure:hero_pairwise} shows results for six representative LLMs.

GPT-5.4 has the broadest win margin (Avg Win 42.9\%, Avg Lose 25.2\%), making it the most consistently fast model.
Kimi-K2.5 follows closely in win rate (39.5\%) with the lowest loss rate (23.1\%), and is rarely the slower solver across opponents,
indicating that its high B$|$P reflects genuine efficiency rather than a favorable task mix.
DeepSeek-V4-Flash falls on the other side, losing more matchups than it wins (33.5\% vs.\ 30.0\%) despite passing more tasks than Kimi. On the 131 tasks where all six models produce correct solutions, the ranking holds, with Kimi-K2.5 (B$|$P, 76.7\%), GPT-5.4 (73.5\%), Claude-Opus-4.6 (70.9\%), DeepSeek-V4-Flash (69.3\%), Qwen3-Coder-Next (58.5\%), and Devstral-2-123B (55.1\%), ruling out correctness-conditioned selection bias.

These patterns consistently hold on the full 16-model matrix (detailed in Appendix~\ref{subsection:appendix_pairwise}), where GPT-5.4 maintains Avg Win 45.9\% and Kimi-K2.5 the lowest Avg Lose (22.2\%).

\textbf{Efficiency leadership rotates across libraries.}
We reveal where each model wins via per-library breakdown. As shown in Table~\ref{table:main_results}, no single model dominates uniformly.
Kimi-K2.5 leads B$|$P on 3 of 5 primary libraries (NumPy 79.6\%, Pandas 74.0\%, SciPy 87.6\%); GPT-5.4-mini leads PyTorch among frontier models (60.9\%); DeepSeek-V4-Flash leads on Polars (72.1\%).
Even within the same family, advantages do not transfer: GPT-5.4 outperforms GPT-5.4-mini on SciPy by 15.0\% (82.3\% vs.\ 67.3\%) yet trails by 7.3\% on PyTorch (53.6\% vs.\ 60.9\%).

These rotating leads and intra-family reversals indicate that efficiency knowledge is library-specific rather than a general capability.
Full per-library breakdowns appear in Appendix~\ref{subsection:library_breakdown_analysis}.

\subsection{Efficiency Gap Diagnosis}
\label{subsection:taxonomy}

For each task, we collect all correct solutions across 16 models and select the maximum-ratio slow--fast pair. We retain only tasks whose slowdown ratio exceeds 1.5$\times$, yielding 868 annotatable pairs.
Two authors independently annotate the pairs and iteratively consolidate labels via a snowball procedure, converging on 5 Level-1 categories and 19 Level-2 sub-patterns (Appendix~\ref{section:appendix_rq2_defs}). Cohen's $\kappa$ is 0.82 (almost perfect agreement~\citep{landis1977measurement}), after which remaining disagreements are resolved through discussion.

\textbf{LLM efficiency deficits are structured.}
The five categories are shown in Table~\ref{table:taxonomy_l1}:
(1) Optimized Primitive Utilization (OPU), using a library's existing optimized primitive instead of re-implementation;
(2) Complexity-Reducing Computation (CRC), reducing computational complexity through algorithmic redesign;
(3) Loop-to-Vectorization (L2V), replacing Python-level iteration with bulk array or table operations;
(4) Data Layout Optimization (DLO), selecting data representations that match the computation and avoid unnecessary intermediates;
and (5) Overhead-Aware Stack Matching (OSM), avoiding unnecessary setup, dispatch, or transfer costs by choosing a lighter computation layer.

\begin{table}[t]
\centering
\small
\renewcommand{\arraystretch}{1.0}
\caption{Five-category efficiency gap taxonomy.}
\label{table:taxonomy_l1}
\begin{tabular}{@{}p{0.69\columnwidth} w{r}{1.15em} w{r}{1.15em} w{r}{1.15em}@{}}
\toprule
\textbf{Category} & \textbf{\%} & \textbf{Med.} & \textbf{P75} \\
\midrule
Optimized Primitive Utilization(OPU) & 23.3 & 5.1 & 18.9 \\
Complexity-Reducing Computation(CRC) & 20.9 & 11.6 & 64.6 \\
Loop-to-Vectorization(L2V) & 19.2 & 10.6 & 36.6 \\
Data Layout Optimization(DLO) & 20.2 & 2.9 & 7.6 \\
Overhead-Aware Stack Matching(OSM) & 16.5 & 14.4 & 98.4 \\
\bottomrule
\end{tabular}
\end{table}

We observe that the five categories differ sharply both in frequency and severity.
OPU is the most frequent (23.3\%), with a moderate median slowdown (5.1$\times$) because a single API substitution typically suffices (\texttt{scipy.spatial.distance.cdist} replacing a triple-nested loop).
OSM is the least frequent (16.5\%) but the most severe (median 14.4$\times$), as unnecessary overhead can dominate execution time by orders of magnitude.
CRC and L2V occupy the middle band (median 11.6$\times$ and 10.6$\times$), requiring algorithmic restructuring or loop elimination; DLO is the mildest (median 2.9$\times$).

\begin{table}[t]
\centering
\scriptsize
\setlength{\tabcolsep}{1.5pt}
\renewcommand{\arraystretch}{1.00}
\definecolor{tblue}{RGB}{232,244,255}
\caption{Category distribution by slow-side model tier.}
\label{table:taxonomy_tier}
\begin{tabular}{@{}l >{\columncolor{tblue}}w{r}{2.0em} w{r}{2.0em} w{r}{2.6em} >{\columncolor{tblue}}w{r}{2.0em} w{r}{2.0em} w{r}{2.6em} >{\columncolor{tblue}}w{r}{2.0em} w{r}{2.0em} w{r}{2.6em}@{}}
\toprule
\textbf{Category} & \multicolumn{3}{c}{\textbf{Closed Frontier}\textit{(191)}} & \multicolumn{3}{c}{\textbf{Open Frontier}\textit{(279)}} & \multicolumn{3}{c}{\textbf{Open Coder}\textit{(398)}} \\
\cmidrule(lr){2-4} \cmidrule(lr){5-7} \cmidrule(lr){8-10}
\rowcolor{white}
& \textbf{\%} & \textbf{Med.} & \textbf{P75} & \textbf{\%} & \textbf{Med.} & \textbf{P75} & \textbf{\%} & \textbf{Med.} & \textbf{P75} \\
\midrule
OPU & \underline{27.2} & 3.0 & 7.6 & \underline{24.4} & 4.2 & 33.8 & 20.6 & 8.8 & 25.2 \\
CRC & 13.6 & 7.4 & 23.4 & 19.0 & 6.7 & 24.4 & \underline{25.6} & 22.1 & 100.7 \\
L2V & 7.3 & 4.9 & 20.1 & 12.5 & 8.3 & 21.1 & \textbf{29.6} & 12.5 & 42.6 \\
DLO & 22.5 & 2.5 & 4.4 & \textbf{26.2} & 2.9 & 11.8 & 14.8 & 4.8 & 7.8 \\
OSM & \textbf{29.3} & 6.2 & 21.0 & 17.9 & 27.2 & 279.7 & 9.3 & 15.5 & 99.0 \\
\bottomrule
\end{tabular}
\end{table}

\textbf{Model tiers exhibit distinct blind spots.}
Table~\ref{table:taxonomy_tier} groups pairs by the slow-side model's tier.
Two categories show clear gradients: L2V rises from 7.3\% (Closed Frontier) to 29.6\% (Open Coder), while OSM drops from 29.3\% to 9.3\%.
Open Coder models still write explicit loops where vectorized calls exist; Closed Frontier models have mastered vectorization but default to heavier computation stacks than the task requires.

Perhaps more surprisingly, Closed Frontier models also show a compositional bias in OPU (27.2\% of their slow-side pairs): they favor constructing solutions from basic operations rather than reusing a library's dedicated primitive, producing correct but suboptimal code.
This build-over-reuse tendency aligns with observations that frontier coding agents prefer custom implementations over adopting existing tools~\citep{amplifying2026claudecode}; our results locate the same pattern at the granularity of DS library APIs and quantify its performance cost.

\begin{figure}[t]
\centering
\includegraphics[width=0.95\columnwidth]{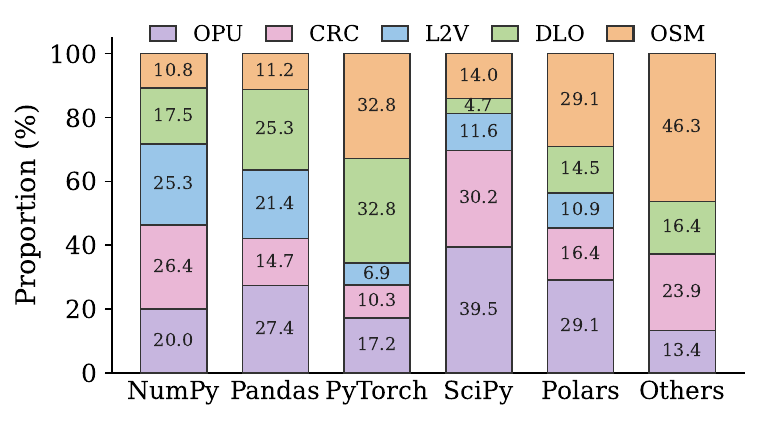}
\caption{Optimization gap distribution (\%) by library.}
\label{figure:taxonomy_lib}
\end{figure}

\textbf{Each library demands a different optimization skill mix.}
Figure~\ref{figure:taxonomy_lib} reveals library-specific profiles.
NumPy concentrates in CRC (26.4\%) and L2V (25.3\%), reflecting the library's emphasis on algorithmic computation and array-level vectorization.
Pandas shifts toward OPU (27.4\%) and DLO (25.3\%): models miss built-in methods (e.g., \texttt{groupby.cummax} instead of \texttt{apply(lambda\ldots)}) and create unnecessary DataFrame copies.
PyTorch and Polars share elevated OSM (32.8\% and 29.1\%), suggesting that models pay framework-dispatch costs disproportionate to the computation.
SciPy is the most OPU-concentrated library (39.5\%), as its rich function inventory means the fastest path is almost always a single specialized call.

These divergent profiles, combined with the tier-specific blind spots above, account for the rotating efficiency rankings in \S\ref{subsection:overall_evaluation}: no model dominates across libraries because each library foregrounds a different optimization category, and each tier struggles with a different subset of those categories.

\subsection{From Diagnosis to Improvement}
\label{subsection:improvement}

We probe the value of our diagnostics via two experiments: taxonomy-guided optimization, and library-conditioned model routing (Lib-Routing).

\subsubsection{Taxonomy-Guided Optimization}
\label{subsubsection:effi_optimization}

\begin{table}[t]
\centering
\scriptsize
\setlength{\tabcolsep}{3.5pt}
\renewcommand{\arraystretch}{1.00}
\definecolor{tblue}{RGB}{232,244,255}
\newcolumntype{B}{>{\columncolor{tblue}}r}
\caption{Exploration of efficiency-aware improvement.}
\label{table:rq3_strategies}
\begin{tabular}{@{}l >{\columncolor{tblue}}w{r}{3em} >{\columncolor{tblue}}w{r}{3em} >{\columncolor{tblue}}w{l}{5.2em} @{\hspace{6pt}{\color{gray!40}\vrule width 0.5pt}\hspace{6pt}} BB B@{}}
\toprule
\rowcolor{white}
\textbf{Strategy} & \textbf{Pass} & \textbf{Beyond} & \textbf{B$|$P} & $\textbf{OPT}_{10}$ & $\textbf{OPT}_{50}$ & \textbf{Regr.} \\
\midrule
\multicolumn{7}{@{}l}{\textit{DeepSeek-V4-Flash \textnormal{($N_{\text{correct}}{=}577$)}}} \\
Basic     & 57.7 & 37.5 & 65.0 & -- & -- & -- \\
EL        & 57.3 & \underline{41.2} & \underline{72.0} {\scriptsize\plus{7.0}} & \underline{22.7} & \underline{11.4} & 3.8 \\
EL w/ Rx  & 56.6 & \textbf{44.1} & \textbf{77.9} {\scriptsize\plus{12.9}} & \textbf{35.5} & \textbf{20.8} & 6.1 \\
\midrule
\multicolumn{7}{@{}l}{\textit{Kimi-K2.5 \textnormal{($N_{\text{correct}}{=}402$)}}} \\
Basic     & 40.2 & 29.6 & 73.6 & -- & -- & -- \\
EL        & 39.4 & \underline{33.7} & \underline{85.5} {\scriptsize\plus{11.9}} & \underline{37.6} & \underline{19.4} & 3.2 \\
EL w/ Rx  & 39.3 & \textbf{34.7} & \textbf{88.3} {\scriptsize\plus{14.7}} & \textbf{54.5} & \textbf{27.6} & 4.5 \\
\bottomrule
\end{tabular}
\end{table}

\textbf{Setup.}
\revise{We test whether the taxonomy from \S\ref{subsection:taxonomy} can serve as actionable guidance for inference-time improvement.
We adopt Effi-Learner~\cite{huang2024effilearner} (EL), a state-of-the-art optimization method: given a correct solution and its line-level execution profile, EL iteratively prompts the model to produce a faster version.
We extend EL with a taxonomy-derived prescription (EL w/ Rx, where Rx denotes prescription): the system prompt is augmented with a library-specific optimization guide listing the five L1 categories, each with its L2 sub-patterns and representative code examples (Appendix~\ref{subsection:appendix_rq3_el_setup}, Figure~\ref{figure:prompt_el_rx}).
We treat the correctness tests as public tests exposed to each method for three rounds of iterative optimization, using them to assess both correctness and efficiency when selecting candidates.
Final evaluation uses the full suite of correctness and stress tests.}

We apply both variants to DeepSeek-V4-Flash ($N_{\text{correct}}{=}577$) and Kimi-K2.5 ($N_{\text{correct}}{=}402$) across 1{,}000 tasks; tasks that fail under Basic prompting retain their original submission.
\revise{$\text{OPT}_{10}$ and $\text{OPT}_{50}$ measure the fraction of originally correct tasks whose optimized solutions remain correct and reduce execution time by at least 10\% and 50\% relative to their Basic-prompting baselines, respectively. Higher values indicate that optimization benefits more tasks. Regression (Regr.)\ measures the fraction of originally correct tasks whose optimized solutions become incorrect or run at least 10\% slower than their baselines.}

\textbf{Taxonomy-guided optimization yields substantial efficiency gains.}
Table~\ref{table:rq3_strategies} reports the results.
Vanilla EL raises B$|$P by +7.0\% (DeepSeek) and +11.9\% (Kimi) at negligible Pass cost (${\leq}$0.8\%), confirming that profile-guided reprompting can improve efficiency without retraining.
Injecting taxonomy-guided optimization amplifies the gain: EL w/ Rx reaches +12.9\% and +14.7\% in B$|$P, producing ${\geq}$2$\times$ speedups on 20.8\% and 27.6\% of originally correct tasks while regression stays within 6.1\%.
\revise{The per-library breakdown (Appendix~\ref{subsection:appendix_rq3_el_setup}) shows that EL w/ Rx outperforms vanilla EL across all six library groups for DeepSeek-V4-Flash and five of six for Kimi-K2.5, indicating that the gains are not concentrated in a single library within DSEffi-Bench.}

\subsubsection{Library-Conditioned Routing}
\label{subsubsection:lib_routing}

\begin{table}[t]
\centering
\small
\setlength{\tabcolsep}{3.92pt}
\caption{Lib-Routing vs.\ closed-frontier baselines on the temporal test set ($N{=}304$).}
\label{table:lib_moe}
\begin{tabular}{@{}l cccc@{}}
\toprule
\textbf{Configuration} & \textbf{Pass} & \textbf{Beyond} & \textbf{B$|$P} & \textbf{\$/task} \\
\midrule
Claude-Opus-4.6             & 63.5 & 43.7 & 68.9 & 0.018 \\
Claude-Opus-4.6 Best@3      & 72.0 & 53.8 & 74.6 & 0.052 \\
GPT-5.4                     & 63.5 & 46.4 & 73.0 & 0.014 \\
GPT-5.4 Best@3              & 73.4 & 58.9 & 80.3 & 0.043 \\
\midrule
\rowcolor{gray!15} Lib-Routing ($K{=}3$)   & \revise{69.4} & \revise{51.4} & \revise{74.1} & 0.004 \\
\bottomrule
\end{tabular}
\end{table}

\textbf{Setup.} Since efficiency knowledge is library-specific, a natural complement to per-model optimization is routing each task to the open model best suited for its library.
\revise{We split the 1{,}000 tasks into training (60\%), validation (10\%), and test (30\%) sets for each library in temporal order.
The training set is used to rank 12 open models and select the top $K{=}3$ experts per library.
The validation set selects $\alpha$ for the composite Pass and B$|$P ranking score, yielding $\alpha{=}0.35$.
At test time, each task is routed to its library's three experts, and the fastest correct solution is submitted.
We compare this fixed configuration against closed-frontier baselines on the unchanged temporal test set (selection and scoring details in Appendix~\ref{subsection:appendix_rq3_moe}).}

\textbf{Library-conditioned routing approaches closed-frontier quality at a fraction of cost.}
\revise{Table~\ref{table:lib_moe} shows that Lib-Routing outperforms both closed-frontier Best@1 baselines in Pass and B$|$P.
It approaches Claude-Opus-4.6 Best@3 in B$|$P (74.1 vs.\ 74.6).
At \$0.004/task, Lib-Routing is 13.0$\times$ cheaper than Claude-Opus-4.6 Best@3.
This result is consistent with the library-level complementarity observed in \S\ref{subsection:taxonomy}, where different models exhibit strengths across different libraries.}

\section{Discussion}
\label{section:discussion}

\textbf{Failure-Type Breakdown of Pass.}
Pass treats both functional errors and solutions exceeding the 5-second timeout as failures, potentially conflating correctness with runtime behavior.
We therefore break down the Pass outcomes by failure type.
Concretely, at T{=}0.0, we classify solutions from all evaluated models as Correct, wrong answer (WA), runtime error (RE), or time limit exceeded (TLE).
Each failed solution is labeled by its first failing test, following competitive-judging practice.
Correctness tests run first, and only passing solutions proceed to stress tests.
The resulting distribution is 45.4\% Correct, 32.9\% WA, 20.4\% RE, and 1.3\% TLE.
The 1.3\% TLE share consists of 0.8\% on correctness tests and 0.5\% on stress tests.
These results show that Pass failures primarily arise from WA and RE rather than timeouts.

\textbf{Future Scoring Protocol for Beyond.}
Beyond is computed relative to the current task-level solution pool, so adding a new model may change the fastest or slowest anchor and rescale existing scores.
To evaluate a new model, researchers generate outputs under the same protocol and run the task-specific tests against the references.
Correct solutions are then time-profiled and added to the corresponding task-level spectra.
Researchers then recompute Beyond for both the new and existing models using the expanded spectra.

\textbf{Licensing and Attribution.}
All SO posts used in this work were published in or after 2021 and are licensed under CC BY-SA 4.0.
Our benchmark retains source links for attribution and distributes SO-derived content under the same license.

\textbf{Societal Impact.}
This work encourages the development of models that generate more efficient DS code, potentially reducing computational costs and energy use in DS workflows.
Our benchmark and taxonomy may support future research on efficiency-aware prompt engineering, model routing, and post-training.

\section{Conclusion}
\label{section:conclusion}

We present DSEffi-Bench, a 1{,}000-instance benchmark that jointly evaluates correctness and execution efficiency of LLM-generated DS code.
Evaluating 16 LLMs across 3 tiers, we find that correctness is not sufficient to characterize efficiency, with notable rank inversions where the models strongest in correctness are not those strongest in efficiency.
A five-category taxonomy traces these deficits to library- and tier-dependent root causes.
The taxonomy further shows potential for guiding single-model optimization and lower-cost cross-model routing within the benchmark.
Together, our benchmark, taxonomy, and diagnostic experiments offer a foundation for future work on improving LLMs' ability to generate efficient DS code.

\section*{Limitations}
\label{section:limitations}

\textbf{Programming Language Scope.}
This work focuses on Python-based DS coding.
Other languages, such as R and MATLAB, are outside our evaluation scope.
LLM-generated code in these languages may exhibit different efficiency patterns and bottlenecks.

\textbf{Environment Dependence.}
Our evaluation protocol relies on test execution, making runtime measurements dependent on hardware resources and software versions.
We therefore evaluate all candidate solutions in the same controlled environment.

\textbf{Data Provenance and Contamination.}
DSEffi-Bench is derived from public Stack Overflow posts.
Evaluated models may have encountered the source content during training, and some recent answers may contain AI-generated or AI-assisted content.
We mitigate these risks through task reformulation and human validation.

\section*{Acknowledgments}
\label{section:acknowledgments}

This work is supported by the National Natural Science Foundation of China under Grant Nos.~62232001 and 62402482. Jie M. Zhang is supported by the GENIUS project (grant no. 23026).

\clearpage
\bibliography{reference}

\clearpage
\appendix
\raggedbottom
\section{Benchmark Details}
\label{section:benchmark_construction_details}

This section provides supplementary details of the benchmark construction described in \S\ref{section:benchmark_construction}.

\subsection{Comparison with Existing Benchmarks}
\label{appendix:related_work_comparison}

Table~\ref{table:benchmark_comparison} positions DSEffi-Bench among 12 related benchmarks along 6 dimensions. Competitive-programming benchmarks (EffiBench(-X), Mercury, COFFE, ENAMEL, EvalPerf)~\cite{huang2024effibench, qing2026effibench, du2024mercury, peng2025coffe, qiu2025efficient, liu2024evaluating} evaluate efficiency on algorithmic problem-solving, anchored to human-curated or model-generated reference solutions. Repository-level benchmarks (SWE-Perf, GSO)~\cite{he2025swe, shetty2026gso} target implementation-level optimization in real-world software projects. Scientific-computing benchmarks (KernelBench, AlgoTune)~\cite{ouyang2025kernelbench, press2026algotune} focus on kernel or numerical solver optimization. None of these address DS library-conditioned efficiency. Existing DS benchmarks (ExeDS, DS-1000, DSCodeBench)~\cite{huang2022execution, lai2023ds, ouyang2026dscodebench} assess only functional correctness without measuring execution efficiency. DSEffi-Bench fills this gap with 1{,}000 human-validated, efficiency-oriented instances spanning 10+ DS libraries, grounded in a multi-model efficiency spectrum plus a human-validated reference for efficiency scoring.

\begin{table*}[t]
\centering
\scriptsize
\setlength{\tabcolsep}{4.0pt}
\caption{Comparison with related benchmarks. The last 3 columns: 1) whether a benchmark evaluates execution efficiency, 2) what type of bottleneck it targets, and 3) what reference anchor grounds the evaluation.}
\label{table:benchmark_comparison}
\begin{tabular}{@{}lrllcll@{}}
\toprule
\textbf{Benchmark} & \textbf{Scale} & \textbf{Task Source} & \textbf{Task Form} & \textbf{Effi.} & \textbf{Effi. Domain} & \textbf{Efficiency Anchor} \\
\midrule
\multicolumn{7}{@{}>{\columncolor{gray!10}}l}{\textit{Competitive Programming}} \\
EffiBench & 1{,}000 & LeetCode & Code Generation & \ding{51} & Algorithmic & 1 human-optimized solution \\
Mercury & 1{,}889 & LeetCode & Code Generation & \ding{51} & Algorithmic & Multiple human solutions \\
COFFE & 756 & HumanEval/MBPP/APPS & Code Generation & \ding{51} & Algorithmic & 1 ground-truth solution \\
ENAMEL & 142 & HumanEval & Code Generation & \ding{51} & Algorithmic & 1 expert-designed solution \\
EvalPerf & 121 & HumanEval/MBPP & Code Generation & \ding{51} & Algorithmic & Multi-model solution clusters \\
\midrule
\multicolumn{7}{@{}>{\columncolor{gray!10}}l}{\textit{Repository-Level Software}} \\
SWE-Perf & 140 & GitHub PRs & Patch Generation & \ding{51} & Impl.-level & 1 expert-optimized PR patch \\
GSO & 102 & GitHub commits & Patch Generation & \ding{51} & Impl.-level & 1 expert-optimized commit \\
\midrule
\multicolumn{7}{@{}>{\columncolor{gray!10}}l}{\textit{Kernels \& Scientific Computing}} \\
KernelBench & 250 & PyTorch ops & Code Optimization & \ding{51} & Hardware & 1 PyTorch reference impl. \\
AlgoTune & 154 & Domain experts & Code Optimization & \ding{51} & Algorithmic & 1 human-authored reference solution \\
\midrule
\multicolumn{7}{@{}>{\columncolor{gray!10}}l}{\textit{Data-Science Code Generation}} \\
ExeDS & 534 & Jupyter notebooks & Code Completion & \ding{55} & --- & --- \\
DS-1000 & 1{,}000 & Stack Overflow & Code Completion & \ding{55} & --- & --- \\
DSCodeBench & 1{,}000 & GitHub & Code Generation & \ding{55} & --- & --- \\
\midrule
\multirow{2}{*}{\textbf{DSEffi-Bench}} & \multirow{2}{*}{\textbf{1{,}000}} & \multirow{2}{*}{\textbf{Stack Overflow}} & \multirow{2}{*}{\textbf{Code Generation}} & \multirow{2}{*}{\ding{51}} & \multirow{2}{*}{\textbf{Library idiom}} & \textbf{Multiple model solutions} \\
 & & & & & & \textbf{+ 1 human-validated reference} \\
\bottomrule
\end{tabular}
\end{table*}

\subsection{Benchmark Instance Schema}
\label{appendix:instance_schema}

Table~\ref{table:instance_schema} lists the fields that constitute a complete instance, divided into the problem description $X$ (provided to models), the evaluation harness $\mathcal{T}$ (used for correctness and efficiency measurement), and the reference solution $S_{\text{ref}}$.

\begin{table*}[t]
\centering
\small
\caption{Benchmark instance schema.}
\label{table:instance_schema}
\begin{tabular}{@{}clp{0.675\textwidth}@{}}
\toprule
 & \textbf{Field} & \textbf{Description} \\
\midrule
\multirow{5}{*}{$X$}
& task\_description & \texttt{str}. Problem statement preserving the original SO post's optimization intent \\
& input\_specification & \texttt{str}. Input types, shapes, and constraints \\
& output\_specification & \texttt{str}. Return type, shape, and semantics \\
& input\_example & \texttt{str}. Concrete input case \\
& output\_example & \texttt{str}. Expected output for the input example \\
\midrule
\multirow{3}{*}{$\mathcal{T}$}
& input\_domain & \texttt{List[str]}. Testing conditions (correctness and stress sub-domains) \\
& input\_generator & \texttt{input\_generator(seed, input\_domain)}. Deterministic generator producing valid inputs \\
& output\_validator & \texttt{output\_validator(pred\_output, ref\_output)}. Correctness checker \\
\midrule
$S_{\text{ref}}$ & reference\_solution & \texttt{def solution(input\_data=None)}. Human-validated solution \\
\bottomrule
\end{tabular}
\end{table*}

\subsection{Efficiency-oriented Post Collection}
\label{appendix:efficiency_oriented_ds_post_collection}

We adopt 21 DS-related tags and 21 efficiency-oriented keywords shown in Figure~\ref{figure:keywords}, surveyed from prior DS benchmarks~\citep{lai2023ds,ouyang2026dscodebench,jing2024dsbench,zhang2025datascibench} and code efficiency studies~\citep{qing2026effibench,qiu2025efficient,waghjale2024ecco,peng2025coffe}. The tag set spans the dominant tools in everyday DS workloads; alias merges (e.g., pandas/dataframe) consolidate semantically equivalent SO tags into a single retrieval channel.

\subsection{Relevance and Feasibility Filtering}
\label{appendix:relevance_and_feasibility_filtering}

This subsection presents the full prompts for the two-stage LLM-as-Judge filtering (\S\ref{subsection:relevance_and_feasibility_filtering}).

\textbf{Question-Level Relevance Filter.}
\label{appendix:prompt_question_relevance_filter}
Given the post title and body, the judge determines whether the post genuinely poses a DS performance optimization problem and returns a structured \texttt{\{is\_relevant, reason\}} decision (Figure~\ref{figure:prompt_q_filter}).

\textbf{Answer-Level Feasibility Filter.}
\label{appendix:prompt_answer_feasibility_filter}
Given the full post (question and accepted answer), the judge determines whether the answer provides a reproducible optimization under our evaluation environment, applying four Feasible criteria and six Non-Feasible exclusions (Figure~\ref{figure:prompt_a_filter}).

\subsection{Instance Transcription}
\label{appendix:task_transcription}

This subsection provides the full prompts for the three transcription steps in \S\ref{subsection:task_transcription}: Problem Description Generation ($X$, Figure~\ref{figure:prompt_problem_gen}), Evaluation Harness Generation ($\mathcal{T}$, Figure~\ref{figure:prompt_harness_gen}), and Multi-Dimensional Scoring (Figure~\ref{figure:prompt_scoring}).

\textbf{Problem Description Generation.}
\label{appendix:prompt_problem_formalization}
Given the original SO question body, the model produces five structured fields (task\_description, input\_specification, output\_specification, input\_example, and output\_example) under the unified function signature \texttt{def solution (input\_data = None)}, with type/shape annotations for all I/O fields.

\textbf{Evaluation Harness Generation.}
\label{appendix:prompt_evaluation_harness_generation}
Given the five $X$-fields and the original SO accepted answer, the model generates the three $\mathcal{T}$-components: \texttt{input\_domain}, \texttt{input\_generator}, and \texttt{output\_validator}.

\textbf{Multi-Dimensional Scoring.}
\label{appendix:prompt_multi_dimensional_scoring}
The quality scorer belongs to a different model family than the transcription model to avoid self-evaluation bias (\S\ref{subsection:task_transcription}). It receives the complete instance and rates it on four dimensions.

\subsection{Human-Validated Reference Solution}
\label{appendix:annotation_statistics}

\textbf{Reference Solution vs.\ Accepted Answer Analysis.} During annotation, we used the SO accepted answer as a starting point and categorized each resulting reference solution into one of three relationships: (1) Faithful Adoption (we directly implement the accepted answer's optimization strategy), (2) Refinement (we follow the same high-level idea but improve implementation details), and (3) Alternative Strategy (we determine that a fundamentally different approach is more efficient).

\begin{figure}[t]
\centering
\includegraphics[width=0.95\linewidth]{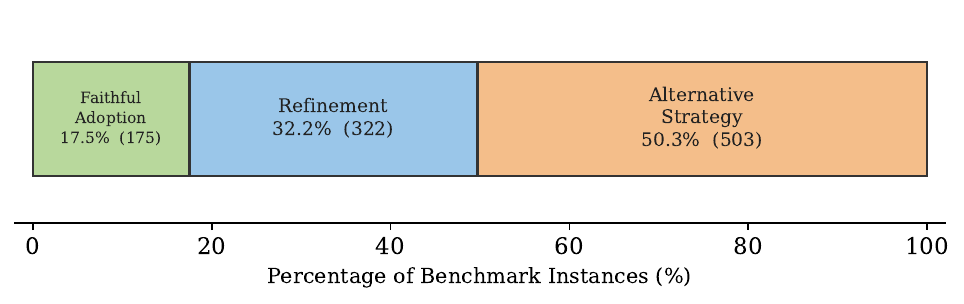}
\caption{Distribution of reference solution strategies relative to the SO accepted answer.}
\label{figure:ref_vs_accepted}
\end{figure}

Figure~\ref{figure:ref_vs_accepted} shows the distribution. About 17.5\% of reference solutions are faithful adoptions of the accepted answer; 32.2\% refine the answer's strategy with implementation improvements; and 50.3\% employ an entirely different optimization approach. 

This distribution reveals several properties of our benchmark:
(1) \textbf{Genuine human effort.} Over 80\% of instances required annotators to go beyond transcribing the SO answer, either improving its implementation or devising a superior strategy within their expertise.
(2) \textbf{Grounded construction.} The accepted answer served as a meaningful starting point: nearly half of all instances (Faithful Adoption and Refinement combined, 49.7\%) build on the answer's core idea, ensuring the reference solutions are anchored in real practitioner knowledge.
\revise{(3) \textbf{Reduced dependence on accepted-answer strategies.} Because 82.5\% of references refine or replace the accepted answer's strategy, most are not direct implementations of the source answer.}

We illustrate each category with a representative example (Figures~\ref{figure:example_faithful}--\ref{figure:example_alternative}). The original SO posts are lightly reformatted (whitespace and HTML markup) for readability.

\noindent\textbf{Example~1 (Faithful Adoption, Figure~\ref{figure:example_faithful}).} Here the reference directly adopts the accepted answer's strategy: extract coordinate index arrays and call \texttt{np.maximum.at} per channel. The only differences are function-signature adaptation (\texttt{solution(input\_data)}) and grid-shape parameterization; the core optimization is identical.

\noindent\textbf{Example~2 (Refinement, Figure~\ref{figure:example_refinement}).} Both solutions use the same mathematical transformation (prefix product plus cumulative sum) to eliminate the Python loop, but the reference streamlines the implementation. The accepted answer constructs an auxiliary array via \texttt{np.insert(x[1:], 0, 0)} and special-cases \texttt{x[0]}; the reference builds \texttt{P} directly (\texttt{P[0]=1}, \texttt{P[1:]=cumprod(d[1:])}), yielding a single expression \texttt{P * cumsum(x / P)} that avoids the intermediate allocation and branch.

\noindent\textbf{Example~3 (Alternative Strategy, Figure~\ref{figure:example_alternative}).} The accepted answer uses Pandas' \texttt{expanding().apply()} with a custom lambda, invoking a Python callback at every row that scans all prior elements, resulting in $O(n^2)$ total work with per-invocation interpreter overhead. After discussion, the annotators adopted a fundamentally different strategy: maintaining a sorted list of prior values and using binary search (\texttt{bisect}) to count elements exceeding the current value in $O(\log k)$ per step (where $k$ is the number of prior elements), while insertion remains $O(k)$ via C-level \texttt{memmove}. The total complexity is still $O(n^2)$, but the elimination of Python-level callbacks and pandas overhead yields a substantial constant-factor speedup.

\noindent\textbf{Scope of the reference.}
The human-validated reference solution is not necessarily the fastest implementation for every task. As our evaluation results confirm (\S\ref{subsection:overall_evaluation}), some LLM-generated solutions outperform the reference on individual instances (the human reference achieves Beyond = 86.6\%, not 100.0\%). However, the reference serves as a grounded correctness and efficiency baseline, informed by both the SO community's domain knowledge and the annotators' optimization expertise.

\subsection{Metric Robustness Analysis}
\label{section:appendix_metric_robustness}

We evaluate whether model rankings are sensitive to the choice of normalization strategy for the spectrum score. Three alternatives are considered:

\begin{itemize}[leftmargin=*,itemsep=2pt]
\item \textbf{Rank-based} (ordinal): $\text{score} = (n - r) / (n - 1)$, where $r$ is the 1-indexed rank by execution time. Ignores execution time magnitudes.
\item \textbf{Linear min-max} (Mercury-style): $\text{score} = (t_{\max} - t_k) / (t_{\max} - t_{\min})$. Uses raw execution time values.
\item \textbf{Log-scale min-max} (adopted): $\text{score} = (\log t_{\max} - \log t_k) / (\log t_{\max} - \log t_{\min})$. Normalizes in log-space.
\end{itemize}

Table~\ref{table:metric_robustness} reports B$|$P under all three strategies. Rankings are robust: the Spearman correlation between log-scale and linear is $\rho = 0.994$ ($p < 10^{-14}$); between log-scale and rank-based is $\rho = 0.961$ ($p < 10^{-8}$). Kendall $\tau$ shows only 2 discordant pairs (out of 120) for linear vs.\ log-scale and 8 for rank-based vs.\ log-scale, all concentrated among models with B$|$P differences below 2\%.

\begin{table}[t]
\centering
\scriptsize 
\setlength{\tabcolsep}{3.2pt}
\renewcommand{\arraystretch}{1.0}
\caption{B$|$P (\%) under three normalization strategies.}
\label{table:metric_robustness}
\begin{tabular}{@{\hspace{4pt}}lrrr@{\hspace{4pt}}}
\toprule
& \multicolumn{3}{c}{\textbf{Normalization Strategy}} \\
\cmidrule(l){2-4}
\textbf{Model} & \textbf{Rank-based} & \textbf{Linear-based} & \textbf{Log-based} \\
\midrule
\multicolumn{4}{>{\columncolor{gray!10}}l}{\textit{Closed Frontier}} \\
\cmidrule[\lightrulewidth]{1-4}
GPT-5.4           & 59.5 & \underline{83.3} & \textbf{71.7} \\
GPT-5.4-mini      & \underline{60.2} & \textbf{83.7} & \underline{71.6} \\
Claude-Opus-4.6   & \textbf{60.9} & 80.2 & 69.2 \\
Claude-Sonnet-4.6 & 52.1 & 79.0 & 66.1 \\
\midrule
\multicolumn{4}{>{\columncolor{gray!10}}l}{\textit{Open Frontier}} \\
\cmidrule[\lightrulewidth]{1-4}
DeepSeek-V4-Flash   & 49.3 & 78.0 & 65.0 \\
DeepSeek-V3.2       & 42.5 & 67.9 & 54.2 \\
Kimi-K2.5           & \textbf{63.6} & \textbf{84.5} & \textbf{73.6} \\
MiniMax-M2.5        & 50.3 & 74.5 & 60.6 \\
Qwen3.5-397B/A17B   & \underline{57.3} & \underline{81.1} & \underline{67.9} \\
\midrule
\multicolumn{4}{>{\columncolor{gray!10}}l}{\textit{Open Coder}} \\
\cmidrule[\lightrulewidth]{1-4}
Devstral-2-123B        & 43.2 & 70.8 & 54.6 \\
Qwen3-Coder-Next       & 43.5 & \textbf{72.5} & \textbf{56.8} \\
Qwen2.5-Coder-32B     & \textbf{46.5} & \underline{71.0} & \underline{54.7} \\
Qwen2.5-Coder-14B     & 38.8 & 63.9 & 46.4 \\
Codestral-22B          & \underline{44.0} & 66.2 & 50.5 \\
DeepSeek-Coder-33B     & 38.8 & 62.1 & 45.5 \\
DeepSeek-Coder-V2-Lite & 35.7 & 57.8 & 42.6 \\
\midrule
\textbf{Human Reference} & 79.0 & 91.9 & 86.6 \\
\bottomrule
\end{tabular}
\end{table}

\noindent\textbf{Why log-scale.}
Per-task execution time spans are wide: the median max/min ratio across tasks with multiple correct solutions is 14.5$\times$, and the 90th percentile reaches 522.0$\times$. Under linear normalization, the median spectrum score among all correct LLM solutions is 0.93 (46.6\% above 0.95), compressing meaningful efficiency differences into a narrow range. Log-scale normalization yields a median of 0.74, preserving discriminative range while assigning equal score intervals to equal multiplicative speedups. This matches how efficiency gains are conventionally interpreted: a 2.0$\times$ speedup carries the same engineering significance whether from 1\,ms to 0.5\,ms or from 1\,s to 0.5\,s.

\subsection{Solution Generation Prompt}
\label{appendix:solution_generation_prompt}

Figure~\ref{figure:prompt_solution} shows the prompt used for all 16~models to generate solutions in DSEffi-Bench. The prompt is intentionally neutral: it asks for a correct solution without explicitly requesting optimization, so that the resulting code reflects each model's natural coding style rather than a prompted optimization effort.

\begin{figure*}[t]
\begin{promptbox}{Example 1 --- Faithful Adoption (SO \#78920216)}
\textbf{Question:} \textit{Optimizing Point Cloud to Voxel Grid with Max Sampling in NumPy}

I have two arrays that represent the point coordinates and values respectively. To max sample from this point cloud, I am initializing a grid with the desired size, and looping over each point to assign the max values:
\begin{Verbatim}[fontsize=\small, breaklines=true]
N = 1000000
coords = np.random.randint(0, 256, size=(N, 3))
vals = np.random.rand(N, 3)

grid = np.zeros((3, 256, 256, 256), dtype=np.float16)
for i, pt in enumerate(coords):
    x, y, z = pt
    grid[0, x, y, z] = max(grid[0, x, y, z], vals[i, 0])
    grid[1, x, y, z] = max(grid[1, x, y, z], vals[i, 1])
    grid[2, x, y, z] = max(grid[2, x, y, z], vals[i, 2])
\end{Verbatim}
Is there a way I can do this through NumPy without the for loop (which is very slow)?

\tcblower

\textbf{Accepted Answer:}

\texttt{np.maximum.at} can be used to perform the max operation in place at specific indices:
\begin{Verbatim}[fontsize=\small, breaklines=true]
indices = coords[:, 0], coords[:, 1], coords[:, 2]
np.maximum.at(grid[0], indices, vals[:, 0])
np.maximum.at(grid[1], indices, vals[:, 1])
np.maximum.at(grid[2], indices, vals[:, 2])
\end{Verbatim}

\textbf{Reference Solution:}
\begin{Verbatim}[fontsize=\small, breaklines=true]
import numpy as np

def solution(input_data=None):
    coords = input_data['coords']
    vals = input_data['vals']
    grid_shape = input_data.get('grid_shape', (3, 256, 256, 256))
    C, X, Y, Z = grid_shape
    grid = np.zeros(grid_shape, dtype=np.float64)
    x = coords[:, 0]
    y = coords[:, 1]
    z = coords[:, 2]
    for c in range(C):                           # same strategy
        np.maximum.at(grid[c], (x, y, z), vals[:, c]) 
    return grid
\end{Verbatim}
\end{promptbox}
\caption{Example 1: Faithful Adoption --- reference directly implements the accepted answer's strategy.}
\label{figure:example_faithful}
\end{figure*}

\begin{figure*}[t]
\begin{promptbox}{Example 2 --- Refinement (SO \#77752955)}
\textbf{Question:} \textit{Dynamically Discounted Cumulative Sum in NumPy}

I have two arrays of the same length: one with values and one with dynamic decay factors; and wish to calculate a vector of the decayed cumulative sum at each position. Using a Python loop to express the desired recurrence:
\begin{Verbatim}[fontsize=\small, breaklines=true]
c = np.empty_like(x)
c[0] = x[0]
for i in range(1, len(x)):
    c[i] = c[i-1] * d[i] + x[i]
\end{Verbatim}
The Python code is very clear but slows things down significantly. Is there a more ``Numpythonic'' way to express this without sacrificing clarity or efficiency?

\tcblower

\textbf{Accepted Answer:}

This is a first-order non-homogeneous recurrence with variable coefficients. Note that \texttt{d[0]} is supposed to be equal to 1. The values in \texttt{c} do not depend on previous values in \texttt{c}:
\begin{Verbatim}[fontsize=\small, breaklines=true]
g = np.insert(x[1:], 0, 0)          # auxiliary shifted array
pd = np.cumprod(d)                   # full prefix product
c = pd * (x[0] + np.cumsum(g / pd)) # x[0] handled separately
\end{Verbatim}

\textbf{Reference Solution:}
\begin{Verbatim}[fontsize=\small, breaklines=true]
import numpy as np

def solution(input_data=None):
    x = np.asarray(input_data['x'], dtype=np.float64)
    d = np.asarray(input_data['d'], dtype=np.float64)
    n = x.shape[0]
    if n == 0:
        return np.array([], dtype=np.float64)
    P = np.empty(n, dtype=np.float64)
    P[0] = 1.0
    if n > 1:
        np.cumprod(d[1:], out=P[1:])  # direct construction, no np.insert
    return P * np.cumsum(x / P)       # unified expression, no x[0] special case
\end{Verbatim}
\end{promptbox}
\caption{Example 2: Refinement --- reference streamlines the accepted answer's approach.}
\label{figure:example_refinement}
\end{figure*}

\begin{figure*}[t]
\begin{promptbox}{Example 3 --- Alternative Strategy (SO \#68057308)}
\textbf{Question:} \textit{Compute Frequency of Numbers with Greater Values Before Current Date}

I'm trying to count the frequency of the number of occurrences a value has been higher than today's value for all dates in the past. Below is a working but inefficient approach:
\begin{Verbatim}[fontsize=\small, breaklines=true]
def freq_greater_than(r):
    smaller_date = df[df.date < r.date]
    larger_num = smaller_date[(smaller_date.num > r.num)]
    return round(len(larger_num) / len(smaller_date) * 100, 2)
\end{Verbatim}
Given a DataFrame with dates and numeric values, I need to compute for each row the percentage of previous dates where the numeric value was greater than the current row's value.

\tcblower

\textbf{Accepted Answer:}

Try with expanding apply + mean over gt comparison:
\begin{Verbatim}[fontsize=\small, breaklines=true]
df['freq_greater_than'] = (
    df['num'].expanding()                     # O(n^2): Python callback at every row
        .apply(lambda s: s.iloc[:-1]             
               .gt(s.iloc[-1]).mean() * 100)  
)
\end{Verbatim}

\textbf{Reference Solution:}
\begin{Verbatim}[fontsize=\small, breaklines=true]
import numpy as np
import pandas as pd
from bisect import bisect_left, bisect_right, insort_left

def solution(input_data=None):
    dates = input_data['dates']
    values = input_data['values']
    n = len(values)
    result = np.full(n, np.nan, dtype=np.float64)
    sorted_prev = []                          # <-- different strategy:
    for i in range(n):                        #     maintain sorted list
        if i > 0:
            pos_right = bisect_right(sorted_prev, values[i])
            count_greater = len(sorted_prev) - pos_right  # O(log n) per step
            result[i] = round(count_greater / len(sorted_prev) * 100, 2)
        insort_left(sorted_prev, values[i])   # insert in O(n), but C-level memmove is fast
    df = pd.DataFrame({
        'date': pd.to_datetime(dates), 'num': values,
        'freq_greater_than': result})
    return df
\end{Verbatim}
\end{promptbox}
\caption{Example 3: Alternative Strategy --- reference departs from the accepted answer entirely.}
\label{figure:example_alternative}
\end{figure*}

\begin{figure*}[t]
\begin{promptbox}{DS Library Tags (with aliases)}
\begin{Verbatim}[fontsize=\small, breaklines=true]
Numerical / scientific:  numpy, scipy, cvxpy
DataFrame / out-of-core: pandas, dataframe, polars, python-polars, dask
Classical ML:            scikit-learn, sklearn, lightgbm, xgboost, catboost
Deep learning:           pytorch, tensorflow, keras, huggingface
Visualization:           matplotlib, seaborn
Compiled acceleration:   cython, numba
\end{Verbatim}
\end{promptbox}

\begin{promptbox}{Efficiency-Related Keywords}
\begin{Verbatim}[fontsize=\small, breaklines=true]
General performance: performance, optimization, optimize, efficiency, efficient, inefficient
Speed:               speed, slow, fast
Profiling:           profile, profiling, bottleneck
Memory:              memory, out of memory
Hardware / parallel: gpu, cuda, parallel
Code patterns:       vectorize, vectorization, loop, overhead
\end{Verbatim}
\end{promptbox}
\caption{DS-related tags and efficiency-oriented keywords used for SO post retrieval.}
\label{figure:keywords}
\end{figure*}

\begin{figure*}[t]
\begin{promptbox}{Question-Level Relevance Filter (System)}
\begin{Verbatim}[fontsize=\small, breaklines=true]
You are a Data Science (DS) expert specialized in performance optimization with Python's DS libraries. Given a Stack Overflow (SO) post, determine if it is relevant to code performance optimization in the DS programming context.
\end{Verbatim}
\end{promptbox}

\begin{promptbox}{Question-Level Relevance Filter (User)}
\begin{Verbatim}[fontsize=\small, breaklines=true]
Evaluate if the following SO post is RELEVANT to Data Science Code Performance Optimization.

**Title**: {title}
**Body**: {body}

### RELEVANT Criteria (Must satisfy ALL):
1. Context: Python code using DS libraries (NumPy, Pandas, Scikit-learn, PyTorch, SciPy, etc.).
2. Goal: Code Performance Optimization (speed, throughput, vectorization, etc. ).
3. Clarity: The post clearly describes the functional scenario, the specific performance optimization requirement, and the input/output data structure.

### IRRELEVANT Criteria (Filter out if ANY applies):
1. Non-Python (R, Julia, Java, C++, SQL, shell).
2. Pure correctness/debugging (unless performance-related).
3. Basics/concept/theory without code optimization.
4. External system/configuration (cluster setup, cloud config).
5. Non-standard input (e.g., images or screenshots).
6. Web/App development (Django, Flask, React, mobile).

### Output Format
{"is_relevant": True or False, "reason": "..."}
\end{Verbatim}
\end{promptbox}
\caption{Prompt for question-level relevance filtering.}
\label{figure:prompt_q_filter}
\end{figure*}

\begin{figure*}[t]
\begin{promptbox}{Answer-Level Feasibility Filter (System)}
\begin{Verbatim}[fontsize=\small, breaklines=true]
You are a Data Science (DS) expert specialized in performance optimization with Python's DS libraries. Given a Stack Overflow (SO) post (Question + Answer), determine if the Answer provides a feasible code performance optimization solution.
\end{Verbatim}
\end{promptbox}

\begin{promptbox}{Answer-Level Feasibility Filter (User)}
\begin{Verbatim}[fontsize=\small, breaklines=true]
Evaluate if the following SO Answer is a FEASIBLE SOLUTION for DS Code Performance Optimization.

**Question Title**: {title}
**Question Body**: {q_body}
**Answer Body**: {a_body}

**Environment**: Linux Ubuntu 22.04, 96 CPU cores, 512 GB RAM, 4 * H200 GPUs, Python 3.12, standard DS libraries.

### Feasibility Criteria (Must satisfy ALL):
1. Reproducibility: concrete solution (code or precise description) that solves the problem.
2. DS Coding: uses Python DS libraries (Numpy, Pandas, etc. ).
3. Optimization: explicitly addresses performance (vectorization, faster API path, algorithmic improvement, etc. ).
4. Clarity: I/O specification can be adopted from the context.

### Non-Feasibility Criteria (Filter out if ANY applies):
1. Negative/impossible ("Cannot be done", "Switch to C++").
2. Non-code solution (change environment, use larger cluster).
3. External dependencies (non-standard setup, external files).
4. External references as primary content (links to docs/tutorials).
5. Non-optimization (only fixes syntax/bug without performance).
6. Irrelevant or off-topic.

### Output Format
{"is_feasible": True or False, "reason": "..."}
\end{Verbatim}
\end{promptbox}
\caption{Prompt for answer-level feasibility filtering.}
\label{figure:prompt_a_filter}
\end{figure*}

\begin{figure*}[t]
\begin{promptbox}{Problem Description Generation (System)}
\begin{Verbatim}[fontsize=\small, breaklines=true]
You are an expert Data Science (DS) coding task designer. Given a seed Stack Overflow (SO) post, transform it into a standard coding task for evaluation.
\end{Verbatim}
\end{promptbox}

\begin{promptbox}{Problem Description Generation (User)}
\begin{Verbatim}[fontsize=\small, breaklines=true]
Analyze the SO post and reformulate it into a standard coding task with: Problem Description, I/O Specification, I/O Examples.

**Question Title**: {title}
**Question Body**: {body}

### Objectives
1. task_description: Transform the SO question into a clear problem in Markdown. Preserve the original questioner's writing style. Minimal edits; keep original intent and constraints.
2. input_specification, output_specification: Assume the solution signature is:
     def solution(input_data):
   Define input as a dictionary with typed keys (e.g., 'x': numpy array of float64 with shape (N,)). Define output with precise type, shape, and semantics.
3. input_example, output_example: Provide a structurally complete dictionary input matching the spec, and the corresponding expected output.

### Output Format (Strict JSON)
{
    "task_description": "...",
    "input_specification": "...",
    "input_example": "...",
    "output_specification": "...",
    "output_example": "..."
}
\end{Verbatim}
\end{promptbox}
\caption{Prompt for problem description generation.}
\label{figure:prompt_problem_gen}
\end{figure*}

\begin{figure*}[t]
\begin{promptbox}{Evaluation Harness Generation (System)}
\begin{Verbatim}[fontsize=\small, breaklines=true]
You are an expert Data Science (DS) coding task designer. You are tasked with designing tests for a Python DS coding problem.
\end{Verbatim}
\end{promptbox}

\begin{promptbox}{Evaluation Harness Generation (User)}
\begin{Verbatim}[fontsize=\small, breaklines=true]
### Task Context
**Problem Description**: {task_desc}
**Input Spec**: {input_spec}
**Output Spec**: {output_spec}
**Input Example**: {input_example}
**Output Example**: {output_example}
**Solution Signature**: def solution(input_data): ...

### Components to Generate
1. input_domain: List of strings. Include correctness tests (prefix correctness-) and stress tests (prefix stress-).
   - correctness-small, correctness-medium, correctness-corner-case
   - stress-medium, stress-large
2. input_generator: def input_generator(seed, input_domain):
   - Set seed via random.seed(seed) and np.random.seed(seed)
   - Branch on input_domain to produce one valid input_data dict
   - Same (seed, domain) -> identical data; different seeds -> different data
   - Yield a single tuple (case_name, input_data)
   - stress inputs must complete within 5s on 96 cores / 512GB RAM
3. output_validator: def output_validator(pred_output, ref_output):
   - Return (True, "") or (False, "reason")
   - Use numpy.testing or np.allclose with tolerance ~1e-4

### Output Format (JSON)
{
    "input_domain": [...],
    "input_generator": "...",
    "output_validator": "..."
}
\end{Verbatim}
\end{promptbox}
\caption{Prompt for evaluation harness generation.}
\label{figure:prompt_harness_gen}
\end{figure*}

\begin{figure*}[t]
\begin{promptbox}{Multi-Dimensional Scoring (System)}
\begin{Verbatim}[fontsize=\small, breaklines=true]
You are DS Code Benchmark Quality Judge. You will receive fields defining ONE benchmark instance for Python DS performance optimization. Evaluate quality and assign scores.
\end{Verbatim}
\end{promptbox}

\begin{promptbox}{Multi-Dimensional Scoring (User)}
\begin{Verbatim}[fontsize=\small, breaklines=true]
## Evaluation Environment
Linux, 96-core CPU, 4x H200, 512GB RAM, no network access.

## Instance Fields
- original_post: {original_post}
- question: {question}
- input_spec: {input_spec}
- output_spec: {output_spec}
- input_example: {input_example}
- output_example: {output_example}
- input_domain: {input_domain}
- input_generator: {input_generator}
- output_validator: {output_validator}

## Hard DISCARD Rules (any triggers -> DISCARD):
A) Not Python DS / not about performance optimization
B) Requires network, special OS features, or non-standard deps
C) Code performs file I/O or modifies environment variables

## Scoring Dimensions (score 1-3)
1) question_clarity: Does the task faithfully reflect the original SO post's intent and DS/performance optimization nature?
2) i_o_clarity: Are input/output specs explicit about types, shapes, constraints? Are examples valid and consistent?
3) input_generator_quality: Does the generator produce valid, diverse inputs with edge cases and sufficient scale?
4) output_validator_quality: Does the validator compare outputs correctly with appropriate numerical tolerances?

Score scale: 3=High quality, 2=Adequate, 1=Low quality (DISCARD)

## Output (JSON only, no markdown)
{
 "analysis": {
    "question_clarity": "...",
    "i_o_clarity": "...",
    "input_generator_quality": "...",
    "output_validator_quality": "..."
    },
 "scores": {
    "question_clarity": N,
    "i_o_clarity": N,
    "input_generator_quality": N,
    "output_validator_quality": N
    },
 "overall_score": N,
 "overall_decision": "KEEP|FIX|DISCARD"
}
\end{Verbatim}
\end{promptbox}
\caption{Prompt for multi-dimensional quality scoring.}
\label{figure:prompt_scoring}
\end{figure*}

\begin{figure*}[t]
\begin{promptbox}{Solution Generation (System)}
\begin{Verbatim}[fontsize=\small, breaklines=true]
You are an expert Data Science (DS) programmer proficient in Python and its ecosystem of DS libraries (NumPy, Pandas, SciPy, PyTorch, etc.). Given a DS coding problem, produce a correct Python solution.
\end{Verbatim}
\end{promptbox}

\begin{promptbox}{Solution Generation (User)}
\begin{Verbatim}[fontsize=\small, breaklines=true]
Solve the following DS coding problem.

## Task
**Task Description**:
{task_description}

**Input Specification**:
{input_specification}

**Output Specification**:
{output_specification}

**Input Example**:
{input_example}

**Output Example**:
{output_example}

## Constraints
1. Define the entry function strictly as: `def solution(input_data=None):`
2. The input is a dictionary; access fields via input_data['key'].
3. Return the result directly (do not print).
4. You may import any standard DS library.

## Output Format
Enclose your solution in a single Python code block:

```python
import ...

def solution(input_data=None):
    # Your implementation
    return result
```
\end{Verbatim}
\end{promptbox}
\caption{Prompt for DS solution generation.}
\label{figure:prompt_solution}
\end{figure*}

\clearpage

\clearpage
\section{Additional Experiment Results}
\label{section:additional_experiment_results}

\subsection{Library Breakdown Analysis}

\label{subsection:library_breakdown_analysis}
\begin{table*}[ht]
\centering
\caption{Per-library results for all 16 models (T{=}0.0).}
\label{table:lib_all_t0}
\renewcommand{\arraystretch}{1.05}
\resizebox{0.95\textwidth}{!}{%
\scriptsize
\setlength{\tabcolsep}{2.5pt}
\begin{tabular}{@{}l rr rr rr rr rr rr rr rr rr rr@{}}
\toprule
 & \multicolumn{2}{c}{\textbf{NumPy}} & \multicolumn{2}{c}{\textbf{Pandas}} & \multicolumn{2}{c}{\textbf{PyTorch}} & \multicolumn{2}{c}{\textbf{SciPy}} & \multicolumn{2}{c}{\textbf{Polars}} & \multicolumn{2}{c}{\textbf{Numba}} & \multicolumn{2}{c}{\textbf{TF}} & \multicolumn{2}{c}{\textbf{Sklearn}} & \multicolumn{2}{c}{\textbf{Dask}} & \multicolumn{2}{c}{\textbf{Others}} \\
 & \multicolumn{2}{c}{\textit{(388)}} & \multicolumn{2}{c}{\textit{(317)}} & \multicolumn{2}{c}{\textit{(68)}} & \multicolumn{2}{c}{\textit{(48)}} & \multicolumn{2}{c}{\textit{(62)}} & \multicolumn{2}{c}{\textit{(25)}} & \multicolumn{2}{c}{\textit{(22)}} & \multicolumn{2}{c}{\textit{(8)}} & \multicolumn{2}{c}{\textit{(10)}} & \multicolumn{2}{c}{\textit{(52)}} \\
\cmidrule(lr){2-3}\cmidrule(lr){4-5}\cmidrule(lr){6-7}\cmidrule(lr){8-9}\cmidrule(lr){10-11}\cmidrule(lr){12-13}\cmidrule(lr){14-15}\cmidrule(lr){16-17}\cmidrule(lr){18-19}\cmidrule(lr){20-21}
\textbf{Model} & \textbf{P} & \textbf{B$|$P} & \textbf{P} & \textbf{B$|$P} & \textbf{P} & \textbf{B$|$P} & \textbf{P} & \textbf{B$|$P} & \textbf{P} & \textbf{B$|$P} & \textbf{P} & \textbf{B$|$P} & \textbf{P} & \textbf{B$|$P} & \textbf{P} & \textbf{B$|$P} & \textbf{P} & \textbf{B$|$P} & \textbf{P} & \textbf{B$|$P} \\
\midrule
\multicolumn{21}{>{\columncolor{gray!10}}l}{\textit{Closed Frontier}} \\
\cmidrule[\lightrulewidth]{1-21}
GPT-5.4 & 76.8 & 78.9 & 65.3 & 71.2 & 64.7 & 53.6 & 64.6 & 82.3 & 72.6 & 59.7 & 48.0 & 31.6 & 9.1 & 43.8 & 25.0 & 44.2 & 20.0 & 0.0 & 50.0 & 60.9 \\
GPT-5.4-mini & 71.4 & 77.2 & 58.0 & 71.9 & 73.5 & 60.9 & 70.8 & 67.3 & 62.9 & 57.4 & 40.0 & 43.3 & 13.6 & 55.6 & 0.0 & -- & 0.0 & -- & 48.1 & 70.3 \\
Claude-Opus-4.6 & 64.9 & 73.4 & 59.6 & 69.6 & 67.6 & 49.4 & 75.0 & 80.3 & 72.6 & 66.9 & 68.0 & 44.4 & 9.1 & 43.7 & 12.5 & 39.6 & 10.0 & 30.0 & 46.2 & 69.0 \\
Claude-Sonnet-4.6 & 67.3 & 68.3 & 57.7 & 69.6 & 55.9 & 48.2 & 62.5 & 74.6 & 67.7 & 57.4 & 60.0 & 50.2 & 9.1 & 92.5 & 0.0 & -- & 20.0 & 6.1 & 46.2 & 60.4 \\
\midrule
\multicolumn{21}{>{\columncolor{gray!10}}l}{\textit{Open Frontier}} \\
\cmidrule[\lightrulewidth]{1-21}
DS-V4-Flash & 65.5 & 70.7 & 54.9 & 63.2 & 58.8 & 40.4 & 68.8 & 67.6 & 46.8 & 72.1 & 60.0 & 24.8 & 13.6 & 87.8 & 0.0 & -- & 10.0 & 30.7 & 53.8 & 69.3 \\
DS-V3.2 & 66.0 & 59.9 & 52.7 & 52.3 & 57.4 & 46.2 & 58.3 & 64.2 & 51.6 & 41.1 & 68.0 & 16.9 & 9.1 & 18.8 & 0.0 & -- & 0.0 & -- & 38.5 & 55.5 \\
Kimi-K2.5 & 44.6 & 79.6 & 41.0 & 74.0 & 41.2 & 59.1 & 27.1 & 87.6 & 53.2 & 58.1 & 24.0 & 45.8 & 9.1 & 93.3 & 0.0 & -- & 0.0 & -- & 32.7 & 61.1 \\
MiniMax-M2.5 & 47.9 & 64.4 & 43.5 & 60.3 & 30.9 & 35.5 & 47.9 & 64.9 & 40.3 & 56.0 & 32.0 & 34.9 & 13.6 & 96.1 & 0.0 & -- & 0.0 & -- & 46.2 & 60.0 \\
Qwen3.5-397B & 67.0 & 70.2 & 57.1 & 72.2 & 52.9 & 51.0 & 43.8 & 69.8 & 58.1 & 62.0 & 56.0 & 34.7 & 9.1 & 94.6 & 12.5 & 40.0 & 10.0 & 68.5 & 46.2 & 60.6 \\
\midrule
\multicolumn{21}{>{\columncolor{gray!10}}l}{\textit{Open Coder}} \\
\cmidrule[\lightrulewidth]{1-21}
Qwen3-Coder-Next & 59.5 & 58.9 & 46.7 & 57.1 & 55.9 & 53.2 & 52.1 & 54.5 & 38.7 & 56.8 & 48.0 & 22.5 & 9.1 & 93.8 & 0.0 & -- & 10.0 & 4.0 & 46.2 & 59.5 \\
Devstral-2-123B & 52.3 & 57.8 & 39.7 & 55.8 & 35.3 & 40.2 & 37.5 & 64.1 & 35.5 & 51.0 & 56.0 & 5.6 & 13.6 & 89.7 & 0.0 & -- & 0.0 & -- & 38.5 & 57.2 \\
Qwen2.5-Coder-32B & 40.7 & 56.5 & 31.5 & 52.2 & 35.3 & 73.3 & 41.7 & 47.4 & 14.5 & 69.9 & 52.0 & 23.6 & 9.1 & 90.1 & 0.0 & -- & 0.0 & -- & 28.8 & 46.3 \\
Qwen2.5-Coder-14B & 39.9 & 44.0 & 29.7 & 50.6 & 33.8 & 53.2 & 43.8 & 39.9 & 11.3 & 61.0 & 48.0 & 29.5 & 0.0 & -- & 0.0 & -- & 0.0 & -- & 30.8 & 51.2 \\
Codestral-22B & 34.5 & 49.5 & 23.0 & 53.5 & 39.7 & 44.0 & 29.2 & 61.1 & 12.9 & 55.2 & 40.0 & 28.1 & 0.0 & -- & 0.0 & -- & 0.0 & -- & 15.4 & 64.8 \\
DS-Coder-33B & 19.3 & 47.2 & 16.1 & 45.2 & 26.5 & 42.4 & 22.9 & 61.7 & 11.3 & 39.5 & 24.0 & 1.8 & 4.5 & 73.9 & 0.0 & -- & 0.0 & -- & 11.5 & 52.8 \\
DS-Coder-V2-Lite & 20.1 & 32.5 & 12.3 & 36.3 & 27.9 & 84.2 & 20.8 & 31.1 & 12.9 & 50.5 & 28.0 & 56.1 & 4.5 & 83.6 & 0.0 & -- & 0.0 & -- & 17.3 & 61.0 \\
\midrule
\textbf{Human Ref.} & 100.0 & 88.5 & 100.0 & 85.8 & 100.0 & 78.7 & 100.0 & 94.2 & 100.0 & 78.1 & 100.0 & 83.8 & 100.0 & 94.4 & 100.0 & 78.0 & 100.0 & 83.5 & 100.0 & 90.4 \\
\bottomrule
\end{tabular}%
}
\end{table*}

Table~\ref{table:lib_all_t0} consolidates Pass and B$|$P for all 16 models across libraries (T=0.0).

NumPy ($N{=}388$) most closely tracks the aggregate results: GPT-5.4 leads in Pass (76.8\%), while Kimi-K2.5 leads B$|$P (79.6\%) despite solving fewer than half the tasks.
Pandas ($N{=}317$) follows a similar pattern, with Kimi-K2.5 leading B$|$P (74.0\%) and Qwen3.5-397B second (72.2\%).
PyTorch ($N{=}68$) stands out: GPT-5.4-mini leads all models in Pass (73.5\%).
Polars ($N{=}62$) presents the sharpest efficiency divergence: DeepSeek-V4-Flash leads B$|$P at 72.1\%, while DeepSeek-V3.2 scores only 41.1\%.
SciPy ($N{=}48$) exhibits particularly strong efficiency, with closed-frontier models performing well (GPT-5.4 82.3\%, Claude-Opus-4.6 80.3\%).
For less-frequent libraries, Numba ($N{=}25$) tasks explicitly require JIT compilation; 
Claude-Opus-4.6 and DeepSeek-V3.2 achieve the highest Pass (68.0\%), whereas the highest B$|$P comes from models that solve fewer tasks, making efficiency comparisons on this small subset more sensitive to the composition of the solved tasks.

\subsection{Pairwise Efficiency Comparison}
\label{subsection:appendix_pairwise}

To isolate efficiency from correctness, we compute pairwise win rates on common-correct subsets. For each ordered pair $(A, B)$, we identify tasks both solve and count the fraction where $A$ is $\geq$10\% faster. This threshold follows the OPT\% metric from PIE~\citep{shypula2023learning}.

\begin{figure*}[t]
\centering
\includegraphics[width=0.9\textwidth]{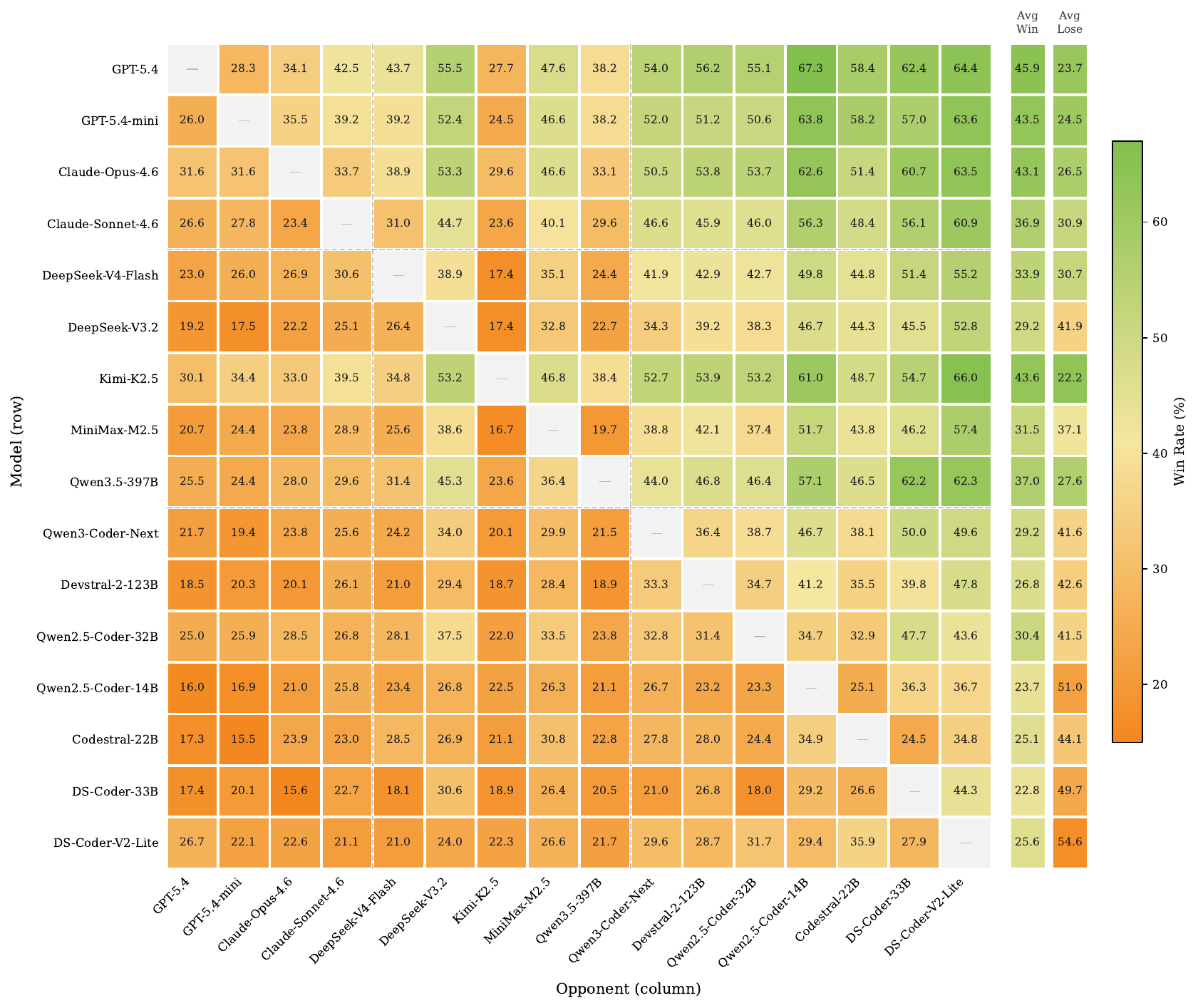}
\caption{Pairwise efficiency win rate (\%) among all 16 models (T{=}0.0). Cell $(A, B)$: fraction of common-correct tasks where $A$ is $\geq$10\% faster. Right columns: task-level average win and lose rates across 15 opponents.}
\label{figure:pairwise_winrate_16model}
\end{figure*}

Figure~\ref{figure:pairwise_winrate_16model} visualizes the full $16 \times 16$ matrix.
GPT-5.4 achieves the highest Avg Win (45.9\%) with the second-lowest Avg Lose (23.7\%), confirming broad efficiency advantage.
Kimi-K2.5 follows closely (43.6\% Win, 22.2\% Lose), matching its strong B$|$P (73.6\% in Table~\ref{table:main_results}).
At the other extreme, DeepSeek-Coder-V2-Lite loses 54.6\% of matchups, and DeepSeek-V3.2 loses 41.9\% despite Pass 56.1\%, reinforcing the correctness--efficiency decoupling in \S\ref{subsection:overall_evaluation}.
Across tiers, all four closed-frontier models achieve Avg Win $>$36\% and Avg Lose $<$31\%; all seven open-coder models show the opposite pattern. The tier-level efficiency gap thus holds consistently at the task level.


\subsection{Efficiency Gap Taxonomy}
\label{section:appendix_rq2}

This section supplements \S\ref{subsection:taxonomy} with the complete two-level taxonomy, category definitions, per-library distribution, and representative examples.
We derive five Level-1 categories and 19 Level-2 sub-patterns from 868 slow--fast code pairs, each naming the optimization strategy the fast solution applies.
Among the five categories, CRC (Complexity-Reducing Computation) captures algorithmic optimizations (e.g., early exit, closed-form substitution, index-based lookup) that parallel the strategies targeted by competitive-programming efficiency benchmarks; the remaining four categories (OPU, L2V, DLO, OSM) require DS library-specific knowledge that those benchmarks do not assess.

\subsubsection{Category Definitions}
\label{section:appendix_rq2_defs}

Table~\ref{table:taxonomy_l2_full} reports the full two-level distribution with median and 75th-percentile slowdowns.

\noindent\textbf{Optimized Primitive Utilization (OPU, 202 pairs).}
The fast solution calls a library's existing optimized primitive instead of reimplementing equivalent logic:
(1) Direct Primitive Call, where an optimized function already exists (e.g., \texttt{np.clip}, \texttt{pd.factorize}); (2) Lightweight API Path, using a faster option within the same function family (e.g., \texttt{transform("min")} vs.\ \texttt{transform(lambda\ldots)}); (3) Thin Dispatch Selection, choosing a lighter dispatch path (e.g., \texttt{scipy.spatial.cKDTree} vs.\ \texttt{KDTree}).

\noindent\textbf{Complexity-Reducing Computation (CRC, 181 pairs).}
The fast solution reduces computational complexity through algorithmic redesign or redundant-work elimination:
(1) Pass Fusion / Early Exit, fusing multiple passes or terminating early; (2) Closed-Form Substitution, replacing a generic numerical routine with an algebraic identity; (3) Index-Based Lookup, replacing brute-force scans with binary search or prefix sums; (4) Purpose-Built Index, introducing a KD-tree, heap, or hash map; (5) Scatter / Segment Reduction, replacing per-group filtering with \texttt{bincount} or \texttt{reduceat}.

\noindent\textbf{Loop-to-Vectorization (L2V, 167 pairs).}
The fast solution replaces Python-level iteration (\texttt{for}/\texttt{while} loops, \texttt{iterrows}, \texttt{apply}) with bulk array or table operations in compiled backends:
(1) Bulk Array Operation, element-wise NumPy/PyTorch vectorization; (2) Table-Native Method, replacing row-wise Pandas/Polars loops with built-in vectorized methods; (3) Shape Broadcasting, leveraging broadcasting or \texttt{einsum} instead of per-element scalar math.

\noindent\textbf{Data Layout Optimization (DLO, 175 pairs).}
The fast solution selects data representations that match the computation, avoiding unnecessary copies or intermediates:
(1) In-Place / View Reuse, reusing buffers, views, or in-place operations; (2) Fit-for-Purpose Container, matching the container to the task (e.g., NumPy array instead of DataFrame for pure array work); (3) Precise Dtype Selection, reducing dtype overhead; (4) Lazy / Chunked Expansion, deferring materialization; (5) Contiguous Layout Alignment, ensuring memory-contiguous layout for cache-friendly access.

\noindent\textbf{Overhead-Aware Stack Matching (OSM, 143 pairs).}
The fast solution avoids unnecessary setup, dispatch, or transfer costs by choosing a lighter computation layer:
(1) Minimal-Path Execution, removing unnecessary validation, copying, or wrapper logic; (2) Native Python for Small Scale, using Python builtins instead of heavy frameworks for small data; (3) Interpreted Sufficiency, avoiding Numba/JIT overhead when NumPy already handles the operation efficiently.

\begin{table}[t]
\centering
\scriptsize
\caption{Two-level taxonomy (868 pairs). Slowdown = slow / fast execution time.}
\label{table:taxonomy_l2_full}
\setlength{\tabcolsep}{4pt}
\resizebox{\columnwidth}{!}{%
\begin{tabular}{@{}lrrrr@{}}
\toprule
\textbf{Level-1 / Level-2}
  & \textbf{Count} & \textbf{\%} & \textbf{Median} & \textbf{p75} \\
\midrule
\rowcolor{gray!10}
\textbf{OPU}
  & \textbf{202} & \textbf{23.3} & \textbf{5.1$\times$} & \textbf{18.9$\times$} \\
\quad Direct Primitive Call
  & 127 & 14.6 & 5.5$\times$ & 21.0$\times$ \\
\quad Lightweight API Path
  & 54 & 6.2 & 4.6$\times$ & 11.4$\times$ \\
\quad Thin Dispatch Selection
  & 21 & 2.4 & 5.5$\times$ & 12.0$\times$ \\
\midrule
\rowcolor{gray!10}
\textbf{CRC}
  & \textbf{181} & \textbf{20.9} & \textbf{11.6$\times$} & \textbf{64.6$\times$} \\
\quad Pass Fusion / Early Exit
  & 79 & 9.1 & 5.1$\times$ & 16.8$\times$ \\
\quad Closed-Form Substitution
  & 33 & 3.8 & 46.7$\times$ & 94.2$\times$ \\
\quad Index-Based Lookup
  & 33 & 3.8 & 33.5$\times$ & 144.7$\times$ \\
\quad Purpose-Built Index
  & 24 & 2.8 & 23.8$\times$ & 43.9$\times$ \\
\quad Scatter / Segment Reduction
  & 12 & 1.4 & 23.2$\times$ & 90.1$\times$ \\
\midrule
\rowcolor{gray!10}
\textbf{L2V}
  & \textbf{167} & \textbf{19.2} & \textbf{10.6$\times$} & \textbf{36.6$\times$} \\
\quad Bulk Array Operation
  & 109 & 12.6 & 12.7$\times$ & 45.2$\times$ \\
\quad Table-Native Method
  & 42 & 4.8 & 7.0$\times$ & 13.3$\times$ \\
\quad Shape Broadcasting
  & 16 & 1.8 & 23.0$\times$ & 67.9$\times$ \\
\midrule
\rowcolor{gray!10}
\textbf{DLO}
  & \textbf{175} & \textbf{20.2} & \textbf{2.9$\times$} & \textbf{7.6$\times$} \\
\quad In-Place / View Reuse
  & 83 & 9.6 & 2.4$\times$ & 6.6$\times$ \\
\quad Fit-for-Purpose Container
  & 62 & 7.1 & 4.7$\times$ & 15.1$\times$ \\
\quad Precise Dtype Selection
  & 17 & 2.0 & 2.4$\times$ & 6.0$\times$ \\
\quad Lazy / Chunked Expansion
  & 7 & 0.8 & 1.8$\times$ & 4.6$\times$ \\
\quad Contiguous Layout Alignment
  & 6 & 0.7 & 10.4$\times$ & 27.6$\times$ \\
\midrule
\rowcolor{gray!10}
\textbf{OSM}
  & \textbf{143} & \textbf{16.5} & \textbf{14.4$\times$} & \textbf{98.4$\times$} \\
\quad Minimal-Path Execution
  & 77 & 8.9 & 14.1$\times$ & 52.4$\times$ \\
\quad Native Python for Small Scale
  & 50 & 5.8 & 8.3$\times$ & 20.1$\times$ \\
\quad Interpreted Sufficiency
  & 16 & 1.8 & 3{,}924.4$\times$ & 14{,}371.9$\times$ \\
\bottomrule
\end{tabular}%
}
\end{table}

\subsubsection{Per-Library Distribution}
\label{section:appendix_rq2_lib}

Table~\ref{table:taxonomy_lib_full} extends the main-paper top-5 cross-tab to all library categories.

\begin{table}[t]
\centering
\small
\caption{Category distribution (\%) by library.}
\label{table:taxonomy_lib_full}
\begin{tabular}{@{}lrrrrrr@{}}
\toprule
\textbf{Library} & $N$ & \textbf{OPU} & \textbf{CRC} & \textbf{L2V} & \textbf{DLO} & \textbf{OSM} \\
\midrule
NumPy & 360 & 20.0 & \textbf{26.4} & 25.3 & 17.5 & 10.8 \\
Pandas & 285 & \textbf{27.4} & 14.7 & 21.4 & 25.3 & 11.2 \\
PyTorch & 58 & 17.2 & 10.3 & 6.9 & \textbf{32.8} & \textbf{32.8} \\
Polars & 55 & \textbf{29.1} & 16.4 & 10.9 & 14.5 & \textbf{29.1} \\
SciPy & 43 & \textbf{39.5} & 30.2 & 11.6 & 4.7 & 14.0 \\
Numba & 22 & 22.7 & 13.6 & 0.0 & 9.1 & \textbf{54.5} \\
Others & 34 & 8.8 & 32.4 & 0.0 & 17.6 & \textbf{41.2} \\
TF & 3 & 0.0 & 33.3 & 0.0 & \textbf{66.7} & 0.0 \\
Matplotlib & 4 & 25.0 & 0.0 & 0.0 & 25.0 & \textbf{50.0} \\
Sklearn & 2 & 0.0 & 50.0 & 0.0 & 0.0 & 50.0 \\
Dask & 2 & 0.0 & 0.0 & 0.0 & 0.0 & \textbf{100.0} \\
\bottomrule
\end{tabular}%
\end{table}

\subsubsection{Model-Level Optimization Profiles}
\label{section:appendix_rq2_model}

Individual models exhibit distinct optimization biases. Table~\ref{table:taxonomy_model_slow} profiles per-model weaknesses (categories where the model most often appears as the slow solution) and Table~\ref{table:taxonomy_model_fast} profiles strengths.

\noindent\textbf{Weakness profiles (Table~\ref{table:taxonomy_model_slow}).}
Each row shows the Level-1 distribution when the model is the slow solution, revealing where it most consistently underperforms.

\begin{table}[t]
\centering
\scriptsize
\setlength{\tabcolsep}{4pt}
\renewcommand{\arraystretch}{1.0}
\caption{Level-1 category distribution (\%) when the model is the slow solution.}
\label{table:taxonomy_model_slow}
\begin{tabular}{@{}lrrrrrrr@{}}
\toprule
\textbf{Model} & $N$ & \textbf{Med.} & \textbf{OPU} & \textbf{CRC} & \textbf{L2V} & \textbf{DLO} & \textbf{OSM} \\
\midrule
\multicolumn{8}{>{\columncolor{gray!10}}l}{\textit{Closed Frontier}} \\
\cmidrule[\lightrulewidth]{1-8}
GPT-5.4            & 55  & 3.0$\times$  & 21.8 &  5.5 &  7.3 & 20.0 & \textbf{45.5} \\
GPT-5.4-mini       & 33  & 5.0$\times$  & \textbf{42.4} & 27.3 &  3.0 & 18.2 &  9.1 \\
Claude-Opus-4.6    & 47  & 3.4$\times$  & 23.4 & 12.8 &  8.5 & \textbf{29.8} & 25.5 \\
Claude-Sonnet-4.6  & 56  & 5.2$\times$  & \textbf{26.8} & 14.3 &  8.9 & 21.4 & 28.6 \\
\midrule
\multicolumn{8}{>{\columncolor{gray!10}}l}{\textit{Open Frontier}} \\
\cmidrule[\lightrulewidth]{1-8}
DS-V4-Flash  & 55  & 4.3$\times$  & 14.5 & 14.5 &  5.5 & \textbf{47.3} & 18.2 \\
DS-V3.2      & 110 & 6.6$\times$  & \textbf{32.7} & 19.1 & 20.9 & 14.5 & 12.7 \\
Kimi-K2.5          & 15  & 6.8$\times$  &  6.7 & \textbf{33.3} &  0.0 & 26.7 & \textbf{33.3} \\
MiniMax-M2.5       & 64  & 11.8$\times$ & 21.9 & 17.2 &  9.4 & 25.0 & \textbf{26.6} \\
Qwen3.5-397B       & 35  & 4.4$\times$  & 25.7 & 22.9 &  8.6 & \textbf{31.4} & 11.4 \\
\midrule
\multicolumn{8}{>{\columncolor{gray!10}}l}{\textit{Open Coder}} \\
\cmidrule[\lightrulewidth]{1-8}
Qwen3-Coder-Next   & 88  & 6.6$\times$  & 21.6 & 22.7 & 18.2 & \textbf{31.8} &  5.7 \\
Devstral-2-123B    & 55  & 11.9$\times$ & 16.4 & \textbf{34.5} & 18.2 & 16.4 & 14.5 \\
Qwen2.5-Coder-32B  & 48  & 7.7$\times$  & 20.8 & \textbf{31.2} & \textbf{31.2} &  6.2 & 10.4 \\
Qwen2.5-Coder-14B  & 76  & 18.9$\times$ & 22.4 & 23.7 & \textbf{36.8} &  6.6 & 10.5 \\
Codestral-22B      & 39  & 7.9$\times$  & 25.6 & 25.6 & \textbf{23.1} & 15.4 & 10.3 \\
DS-Coder-33B & 34  & 13.9$\times$ & 14.7 & 32.4 & \textbf{35.3} & 11.8 &  5.9 \\
DS-Coder-V2-Lite   & 58  & 25.6$\times$ & 20.7 & 15.5 & \textbf{48.3} &  6.9 &  8.6 \\
\bottomrule
\end{tabular}
\end{table}

The most striking pattern is an optimization-weakness migration across capability levels: the dominant weakness shifts from Loop-to-Vectorization (48.3\% for DS-Coder-V2-Lite) toward Overhead-Aware Stack Matching (45.5\% for GPT-5.4).
Frontier models have largely mastered basic vectorization but develop a new blind spot: over-engineering.
GPT-5.4, the best model by Pass rate, has the lowest median slowdown (3.0$\times$), yet 45.5\% of its slow pairs are OSM---importing heavy frameworks, adding unnecessary validation, or wrapping already-vectorized code in defensive scaffolding.
DLO emerges as the dominant weakness for several otherwise strong models: DeepSeek-V4-Flash (47.3\%), Qwen3-Coder-Next (31.8\%), and Qwen3.5-397B (31.4\%). This suggests that data layout awareness is a distinct skill from algorithmic or API knowledge.

\noindent\textbf{Strength profiles (Table~\ref{table:taxonomy_model_fast}).}
When a model appears as the fast solution, the category distribution reveals which optimization strategies it successfully applies.

\begin{table}[t]
\centering
\scriptsize
\setlength{\tabcolsep}{3.5pt}
\renewcommand{\arraystretch}{1.0}
\caption{Level-1 category distribution (\%) when the model is the fast solution. Models with $N{<}20$ are omitted.}
\label{table:taxonomy_model_fast}
\begin{tabular}{@{}lrrrrrrr@{}}
\toprule
\textbf{Model} & $N$ & \textbf{Med.} & \textbf{OPU} & \textbf{CRC} & \textbf{L2V} & \textbf{DLO} & \textbf{OSM} \\
\midrule
\multicolumn{8}{>{\columncolor{gray!10}}l}{\textit{Closed Frontier}} \\
\cmidrule[\lightrulewidth]{1-8}
GPT-5.4            & 130 & 8.7$\times$  & 21.5 & \textbf{30.8} & 23.1 & 13.8 & 10.8 \\
GPT-5.4-mini       & 105 & 6.2$\times$  & 19.0 & 21.9 & 20.0 & 22.9 & 16.2 \\
Claude-Opus-4.6    & 105 & 10.8$\times$ & 21.9 & \textbf{24.8} & 18.1 & 20.0 & 15.2 \\
Claude-Sonnet-4.6  & 64  & 5.6$\times$  & \textbf{25.0} & 21.9 & 18.8 & \textbf{25.0} &  9.4 \\
\midrule
\multicolumn{8}{>{\columncolor{gray!10}}l}{\textit{Open Frontier}} \\
\cmidrule[\lightrulewidth]{1-8}
Kimi-K2.5          & 78  & 6.8$\times$  & \textbf{29.5} & 25.6 & 21.8 & 17.9 &  5.1 \\
Qwen3.5-397B       & 71  & 5.9$\times$  & 23.9 & 15.5 & \textbf{28.2} & 19.7 & 12.7 \\
DS-V3.2      & 50  & 5.8$\times$  & 14.0 & 14.0 & 24.0 & \textbf{26.0} & 22.0 \\
DS-V4-Flash  & 45  & 8.4$\times$  & \textbf{33.3} & 17.8 & 17.8 & 11.1 & 20.0 \\
MiniMax-M2.5       & 42  & 6.5$\times$  & \textbf{31.0} & 14.3 & 19.0 & 16.7 & 19.0 \\
\midrule
\multicolumn{8}{>{\columncolor{gray!10}}l}{\textit{Open Coder}} \\
\cmidrule[\lightrulewidth]{1-8}
Qwen3-Coder-Next   & 43  & 5.7$\times$  & 18.6 & 23.3 & 11.6 & 18.6 & \textbf{27.9} \\
Devstral-2-123B    & 24  & 4.7$\times$  & \textbf{37.5} &  8.3 & 16.7 & 20.8 & 16.7 \\
Qwen2.5-Coder-32B  & 42  & 4.4$\times$  & 19.0 & 14.3 &  7.1 & 21.4 & \textbf{38.1} \\
Qwen2.5-Coder-14B  & 26  & 9.0$\times$  & 23.1 & 11.5 & 11.5 & \textbf{26.9} & \textbf{26.9} \\
\bottomrule
\end{tabular}
\end{table}

The fast-side profiles complement the weakness view.
Closed Frontier models spread their optimization strength broadly, with GPT-5.4 strongest in CRC (30.8\%) and Claude-Opus-4.6 similarly CRC-leaning (24.8\%), reflecting proficiency in algorithmic redesign.
Kimi-K2.5’s fast-side profile is OPU-heavy (29.5\%) and has the lowest OSM share (5.1\%), consistent with its preference for direct, lightweight code paths.
Open Coder models, when they do appear on the fast side, concentrate in OSM (Qwen2.5-Coder-32B at 38.1\%, Qwen3-Coder-Next at 27.9\%): their ``wins'' often come from writing simpler code that avoids overhead, rather than applying sophisticated optimizations.

Orthogonal to the per-category profiles, JIT compilation overhead is a recurring cross-model pattern.
Of the 868 pairs, 49 (5.6\%) contain Numba or JIT decorators in the slow code; their median slowdown is 1{,}131.0$\times$ versus 6.5$\times$ for non-JIT pairs.
DeepSeek-V3.2 uses Numba in 14.4\% of its slow pairs, DeepSeek-V4-Flash in 12.7\%.
On the small-to-medium data sizes typical of DS tasks, JIT overhead can exceed the computation it was meant to accelerate.

\subsubsection{Case Study}
\label{section:appendix_rq2_cases}

Below we present one slow--fast pair per Level-1 category.
Task descriptions and I/O specifications are condensed for brevity.


\textbf{Optimized Primitive: triple nested loop vs.\ \texttt{scipy.spatial.distance.cdist} (389.4$\times$).}
DeepSeek-V3.2 writes a triple-nested Python loop over points and dimensions, producing $O(NMD)$ interpreter iterations.
Claude-Sonnet-4.6 recognizes that \texttt{scipy.spatial.distance.cdist} performs the identical computation in compiled C, eliminating all Python-level iteration (Figure~\ref{figure:case_opu}).



\textbf{Complexity-Reducing: per-matrix \texttt{np.linalg.det} loop vs.\ closed-form $ad{-}bc$ (183.8$\times$).}
Devstral-2-123B calls \texttt{np.linalg.det} in a Python list comprehension over $N$ matrices, invoking LAPACK LU decomposition per element.
Claude-Opus-4.6 exploits the $2{\times}2$ structure: $\det = ad - bc$ is four element-wise array operations, fully vectorized across the batch (Figure~\ref{figure:case_crc}).



\textbf{Loop-to-Vec: nested loop with per-cell \texttt{np.sum} vs.\ nine-slice vectorization (167.4$\times$).}
The nested Python loop calls \texttt{np.sum} on a $3{\times}3$ slice per pixel, incurring function-call overhead proportional to the image area.
The vectorized version sums nine shifted array views in a single expression, eliminating Python iteration entirely in favor of compiled NumPy (Figure~\ref{figure:case_l2v}).



\textbf{Data Layout: DataFrame-based row replication vs.\ pre-allocated NumPy arrays (95.5$\times$).}
GPT-5.4-mini stays in the DataFrame layer throughout: \texttt{index.repeat} expands the index, then \texttt{pd.concat} materializes $n \cdot R$ copies of the row, and a boolean mask overwrites positions, triggering multiple internal copies.
DeepSeek-V4-Flash drops to NumPy arrays: pre-allocated buffers filled with the repeated value, then original rows scattered into place via computed indices, constructing the DataFrame only once at the end (Figure~\ref{figure:case_dlo}).

\vspace{1em}

\textbf{Stack Matching: function-level \texttt{import} and dead-code branches (218.4$\times$).}
Both models write the same core computation: \texttt{(A * B).sum(dim=1)}.
The 218.4$\times$ gap comes entirely from GPT-5.4's defensive scaffolding: \texttt{import torch} inside the function body (re-executing module lookup on every call) and a dead \texttt{if input\_data is None} branch that constructs dummy tensors.
The fast code hoists the import to module scope and removes the dead branch.
This is a pure overhead example: the algorithm is identical, but the ``production-ready'' wrapper dominates execution time (Figure~\ref{figure:case_osm}).

\subsection{Memorization Analysis}
\label{subsection:appendix_memorization}

\revise{
To assess potential memorization, we split the 1{,}000 tasks by whether the source Stack Overflow question was created before or after June 2024, following the temporal split methodology of LiveCodeBench~\citep{jain2025livecodebench}. This yields 904 pre-cutoff and 96 post-cutoff tasks (Table~\ref{table:cutoff}). Because the post-cutoff subset is limited in size, we interpret this analysis as a sensitivity check rather than conclusive evidence against memorization.
}

\begin{table}[t]
\centering
\scriptsize
\setlength{\tabcolsep}{3.125pt}
\renewcommand{\arraystretch}{1.0}
\caption{Pre- vs.\ post-cutoff results. Cutoff: June 2024.}
\label{table:cutoff}
\begin{tabular}{@{}l rrr rrr@{}}
\toprule
 & \multicolumn{3}{c}{\textbf{Pre}} & \multicolumn{3}{c}{\textbf{Post}} \\
\cmidrule(lr){2-4}\cmidrule(lr){5-7}
\textbf{Model} & \textbf{P} & \textbf{B} & \textbf{B$|$P} & \textbf{P} & \textbf{B} & \textbf{B$|$P} \\
\midrule
\multicolumn{7}{>{\columncolor{gray!10}}l}{\textit{Closed Frontier}} \\
\cmidrule[\lightrulewidth]{1-7}
GPT-5.4       & 67.6 & 48.2 & 71.4 & 60.4 & 45.7 & 75.7 \\
GPT-5.4-mini  & 62.1 & 44.7 & 72.0 & 63.5 & 43.1 & 67.9 \\
Claude-Opus-4.6 & 61.4 & 42.2 & 68.7 & 60.4 & 44.4 & 73.6 \\
Claude-Sonnet-4.6 & 59.3 & 39.2 & 66.1 & 63.5 & 42.0 & 66.1 \\
\midrule
\multicolumn{7}{>{\columncolor{gray!10}}l}{\textit{Open Frontier}} \\
\cmidrule[\lightrulewidth]{1-7}
DS-V4-Flash & 57.9 & 37.4 & 64.7 & 56.2 & 38.0 & 67.5 \\
DS-V3.2 & 56.1 & 30.7 & 54.7 & 56.2 & 27.7 & 49.3 \\
Kimi-K2.5 & 40.5 & 30.2 & 74.5 & 37.5 & 24.2 & 64.6 \\
MiniMax-M2.5 & 42.5 & 25.7 & 60.5 & 45.8 & 28.3 & 61.7 \\
Qwen3.5-397B & 58.0 & 39.4 & 68.0 & 54.2 & 36.2 & 66.8 \\
\midrule
\multicolumn{7}{>{\columncolor{gray!10}}l}{\textit{Open Coder}} \\
\cmidrule[\lightrulewidth]{1-7}
Qwen3-Coder-Next & 51.8 & 29.4 & 56.8 & 38.5 & 21.9 & 56.7 \\
Devstral-2-123B & 43.6 & 23.9 & 54.8 & 37.5 & 19.7 & 52.6 \\
Qwen2.5-Coder-32B & 35.5 & 19.4 & 54.6 & 20.8 & 11.8 & 56.7 \\
Qwen2.5-Coder-14B & 33.6 & 15.5 & 46.1 & 25.0 & 12.9 & 51.4 \\
Codestral-22B & 27.5 & 13.8 & 50.2 & 26.0 & 13.9 & 53.2 \\
DS-Coder-33B & 17.5 & 7.9 & 45.0 & 17.7 & 8.9 & 50.0 \\
DS-Coder-V2-Lite & 17.5 & 7.7 & 43.8 & 13.5 & 3.9 & 28.7 \\
\midrule
\textbf{Human Reference} & 100.0 & 86.8 & 86.8 & 100.0 & 84.1 & 84.1 \\
\bottomrule
\end{tabular}
\end{table}

\revise{
Pass rates drop slightly on post-cutoff tasks (mean $-$3.5\%).
B$|$P does not decline systematically: 8 of 16 models score higher on post-cutoff tasks, and rankings are well-preserved (Spearman $\rho = 0.87$ for B$|$P, $0.95$ for Pass).
These results are consistent with the main correctness--efficiency decoupling, but do not eliminate contamination concerns because all source posts remain public and the post-cutoff subset is relatively small.
}

\subsection{From Diagnosis to Remedy: Experimental Details}
\label{subsection:appendix_rq3}

This section supplements \S\ref{subsection:improvement} with method details, experimental protocols, and breakdowns for both experiments.

\subsubsection{Taxonomy-Guided Optimization}
\label{subsection:appendix_rq3_el_setup}

\begin{table}[t]
\centering
\scriptsize
\setlength{\tabcolsep}{2.5pt}
\renewcommand{\arraystretch}{1.00}
\caption{Per-library B$|$P (\%) on Basic-correct tasks. $N$: tasks correct under Basic in each library.}
\label{table:el_per_lib}
\begin{tabular}{@{}l r rr rr @{\hspace{8pt}{\color{gray!40}\vrule width 0.4pt}\hspace{8pt}} r rr rr@{}}
\toprule
& \multicolumn{5}{c}{\textbf{DeepSeek-V4-Flash}} & \multicolumn{5}{c}{\textbf{Kimi-K2.5}} \\
\cmidrule(lr){2-6} \cmidrule(l){7-11}
\textbf{Library} & $N$ & \textbf{EL} & $\Delta$ & \textbf{Rx} & $\Delta$ & $N$ & \textbf{EL} & $\Delta$ & \textbf{Rx} & $\Delta$ \\
\midrule
NumPy   & 254 & 77.8 & \plus{7.1}  & 82.6 & \plus{12.0} & 173 & 88.9 & \plus{9.3}  & 90.4 & \plus{10.8} \\
Pandas  & 174 & 74.2 & \plus{11.0} & 76.9 & \plus{13.7} & 130 & 90.6 & \plus{16.6} & 92.8 & \plus{18.8} \\
PyTorch &  40 & 47.3 & \plus{6.9}  & 51.8 & \plus{11.4} &  28 & 73.0 & \plus{13.9} & 71.3 & \plus{12.2} \\
SciPy   &  33 & 76.8 & \plus{9.2}  & 85.6 & \plus{18.0} &  13 & 93.5 & \plus{5.9}  & 97.2 & \plus{9.6}  \\
Polars  &  29 & 75.6 & \plus{3.5}  & 78.1 & \plus{6.0}  &  33 & 78.1 & \plus{20.0} & 83.2 & \plus{25.1} \\
Others  &  47 & 62.6 & \plus{7.1}  & 78.4 & \plus{23.0} &  25 & 82.9 & \plus{22.9} & 88.6 & \plus{28.6} \\
\midrule
\textbf{All} & \textbf{577} & \textbf{73.2} & \plus{8.3} & \textbf{78.5} & \plus{13.5} & \textbf{402} & \textbf{87.4} & \plus{13.8} & \textbf{89.5} & \plus{15.9} \\
\bottomrule
\end{tabular}
\end{table}

\noindent\textbf{Method.}
Effi-Learner (EL) operates in an iterative feedback loop.
In each round, the system (1)~executes the current best code on the benchmark's correctness-test inputs and collects a line-level execution time profile, (2)~constructs a prompt containing the original task description, the current code, and the profile highlighting hotspot lines, and (3)~sends the prompt to the same base model requesting a faster implementation.
If the returned code passes all correctness tests and runs faster than the current best on those same tests, it replaces it; otherwise the previous version is carried forward.
Each successive round receives the output of the previous round (not the original Basic code) along with an updated profile, enabling incremental optimization.
EL w/ Rx augments the EL prompt with a taxonomy-derived optimization prescription listing the five L1 categories, each with its L2 sub-patterns and one-line code examples.
This prescription is prepended so the model can match profiled hotspots to known optimization strategies before generating a replacement; the iteration mechanism is otherwise identical to vanilla EL (Figure~\ref{figure:prompt_el_rx}).

\noindent\textbf{Experimental protocol.}
We apply both EL and EL w/ Rx to DeepSeek-V4-Flash ($N_{\text{correct}}{=}577$) and Kimi-K2.5 ($N_{\text{correct}}{=}402$), prompting at T{=}0.0.
Only Basic-correct tasks are submitted for optimization; Basic-failed tasks retain their original submission and are never reprompted.
We set the iteration budget to 3 rounds per task; the per-round breakdown (Table~\ref{table:el_per_round}) confirms this is sufficient, with accepted improvements concentrated in the first round and diminishing returns after.
The final submission is evaluated on the full test suite; if it fails correctness or degrades execution time by ${\geq}$10\%, it counts as a regression.

\noindent\textbf{Per-library results.}
Table~\ref{table:el_per_lib} reports B$|$P on the Basic-correct subset per library for both models.
Unlike Table~\ref{table:rq3_strategies}, which reports full-benchmark outcomes after optimization, this breakdown fixes the Basic-correct subset to isolate efficiency changes within each library.
EL w/ Rx delivers uniformly larger gains than vanilla EL across nearly all library groups and both models.

The variation across libraries is substantial and model-dependent.
For DeepSeek-V4-Flash, Others sees the largest Rx gain (+23.0\%) and SciPy follows (+18.0\%), consistent with their OSM- and OPU-concentrated taxonomy profiles where the prescription directly guides the model toward lighter execution stacks or specialized primitives.
For Kimi-K2.5, Others (+28.6\%) and Polars (+25.1\%) benefit most, reflecting Kimi's stronger baseline algorithmic choices that leave framework-level overhead as the primary remaining gap.
PyTorch is the only library where Rx occasionally underperforms EL (Kimi: +12.2 vs.\ +13.9), suggesting that knowledge-augmented reprompting can destabilize solutions in frameworks with complex dispatch semantics.

Across both models, EL retains ${\geq}$98\% of Basic-correct tasks (DeepSeek: 573/577; Kimi: 394/402), and EL w/ Rx retains ${\geq}$97\% (DeepSeek: 566/577; Kimi: 393/402), confirming that efficiency-focused reprompting rarely breaks correctness.

\noindent\textbf{Per-round results.}
Table~\ref{table:el_per_round} reports how many tasks are newly optimized in each round.
R1 captures 69--75\% of all accepted improvements; by R3, at most 32 tasks are newly accepted in any setting, justifying the three-round budget.

\begin{table}[t]
\centering
\small
\caption{New optimizations accepted per round.}
\label{table:el_per_round}
\begin{tabular}{@{}ll ccc@{}}
\toprule
\textbf{Model} & \textbf{Method} & \textbf{R-1} & \textbf{R-2} & \textbf{R-3} \\
\midrule
\multirow{2}{*}{DS-V4-Flash}
  & EL       & 155 & 46 & 25 \\
  & EL w/ Rx & 201 & 60 & 32 \\
\midrule
\multirow{2}{*}{Kimi-K2.5}
  & EL       & 159 & 42 & 10 \\
  & EL w/ Rx & 198 & 51 & 17 \\
\bottomrule
\end{tabular}
\end{table}

\subsubsection{Library-Conditioned Routing}
\label{subsection:appendix_rq3_moe}

\revise{\noindent\textbf{Method.}
We split the 1{,}000 benchmark instances into training (60\%), validation (10\%), and test (30\%) sets via a per-library temporal split.
Within each library, instances are sorted by SO creation date.
The earliest 60\% form the training set, the next 10\% form the validation set, and the latest 30\% form the test set.
This yields 597 training, 99 validation, and 304 test instances, while preserving the original temporal test set.}

\revise{\noindent\textbf{Expert and hyperparameter selection.}
On the training set, we compute per-library Pass and B$|$P (T{=}0.0) for all 12 open-weight models (5 Open Frontier + 7 Open Coder).
Models are ranked independently by Pass and B$|$P within each library.
We combine the rankings using $\alpha \cdot \text{rank}_{\text{Pass}} + (1{-}\alpha) \cdot \text{rank}_{\text{B}|\text{P}}$ and designate the top $K{=}3$ models as experts.
We then sweep $\alpha \in [0, 1]$ at 0.05 increments on the validation set.
The validation set selects $\alpha{=}0.35$, which achieves the highest B$|$P (77.4), with Beyond used to break ties.
After fixing $\alpha$ and the corresponding experts, we evaluate the routing configuration on the test set.}

\revise{\noindent\textbf{Per-library expert assignment.}
Table~\ref{table:moe_routing} shows the experts selected at $\alpha{=}0.35$.
NumPy, Pandas, Polars, and Others share the same three experts, although their ranking differs.
PyTorch and SciPy select different experts, reflecting variation in model strengths across libraries.}

\begin{table}[H]
\centering
\small
\setlength{\tabcolsep}{2pt}
\caption{\revise{Per-library experts selected from the training set using the validation-selected $\alpha{=}0.35$ ($K{=}3$).}}
\label{table:moe_routing}
\begin{tabular}{@{}l lll@{}}
\toprule
\textbf{Library} & \textbf{Expert 1} & \textbf{Expert 2} & \textbf{Expert 3} \\
\midrule
NumPy   & Qwen3.5-397B  & DS-V4-Flash   & Kimi-K2.5 \\
Pandas  & Qwen3.5-397B  & Kimi-K2.5     & DS-V4-Flash \\
PyTorch & \revise{Qwen3.5-397B} & \revise{Qwen3-C-Next} & Qwen2.5-C-32B \\
SciPy   & \revise{DS-V4-Flash}  & \revise{DS-V3.2}   & \revise{MiniMax-M2.5} \\
Polars  & Kimi-K2.5     & DS-V4-Flash   & Qwen3.5-397B \\
Others  & \revise{Qwen3.5-397B} & \revise{DS-V4-Flash} & Kimi-K2.5 \\
\bottomrule
\end{tabular}
\end{table}


\begin{figure*}[t]
\begin{promptbox}{Q\#69531729 $\cdot$ NumPy $\cdot$ 389.4$\times$ slowdown}
\textbf{Task.} Compute pairwise Euclidean distances between two point sets.\\[2pt]
\textbf{Input.} \texttt{'XA'}: float64 array $(N, D)$; \texttt{'XB'}: float64 array $(M, D)$.\\
\textbf{Output.} Float64 array $(N, M)$ where element $(i,j)$ is the Euclidean distance between \texttt{XA[i]} and \texttt{XB[j]}.
\end{promptbox}

\begin{promptbox}{Slow (DeepSeek-V3.2)}
\begin{Verbatim}[fontsize=\small, breaklines=true]
import numpy as np

def solution(input_data=None):
    XA = input_data['XA']
    XB = input_data['XB']
    N, D = XA.shape
    M, _ = XB.shape
    distances = np.zeros((N, M), dtype=np.float64)
    for i in range(N):
        for j in range(M):
            sq_dist = 0.0
            for k in range(D):
                diff = XA[i, k] - XB[j, k]
                sq_dist += diff * diff
            distances[i, j] = np.sqrt(sq_dist)
    return distances
\end{Verbatim}
\end{promptbox}

\begin{promptbox}{Fast (Claude-Sonnet-4.6)}
\begin{Verbatim}[fontsize=\small, breaklines=true]
import numpy as np
from scipy.spatial.distance import cdist

def solution(input_data=None):
    XA = np.array(input_data['XA'], dtype=np.float64)
    XB = np.array(input_data['XB'], dtype=np.float64)
    return cdist(XA, XB, metric='euclidean')
\end{Verbatim}
\end{promptbox}
\caption{Optimized Primitive Utilization: triple nested loop vs.\ \texttt{cdist} (389.4$\times$).}
\label{figure:case_opu}
\end{figure*}

\begin{figure*}[t]
\begin{promptbox}{Q\#78368046 $\cdot$ NumPy $\cdot$ 183.8$\times$ slowdown}
\textbf{Task.} Compute determinants of a batch of $2{\times}2$ matrices.\\[2pt]
\textbf{Input.} \texttt{'matrix\_batch'}: float64 array $(N, 2, 2)$.\\
\textbf{Output.} Float64 array $(N,)$ where element $i$ is the determinant of \texttt{matrix\_batch[i]}.
\end{promptbox}

\begin{promptbox}{Slow (Devstral-2-123B)}
\begin{Verbatim}[fontsize=\small, breaklines=true]
import numpy as np

def solution(input_data=None):
    matrix_batch = input_data['matrix_batch']
    determinants = np.array(
        [np.linalg.det(matrix) for matrix in matrix_batch])
    return determinants
\end{Verbatim}
\end{promptbox}

\begin{promptbox}{Fast (Claude-Opus-4.6)}
\begin{Verbatim}[fontsize=\small, breaklines=true]
import numpy as np

def solution(input_data=None):
    matrix_batch = input_data['matrix_batch']
    return (matrix_batch[:, 0, 0] * matrix_batch[:, 1, 1]
          - matrix_batch[:, 0, 1] * matrix_batch[:, 1, 0])
\end{Verbatim}
\end{promptbox}
\caption{Complexity-Reducing Computation: \texttt{np.linalg.det} loop vs.\ closed-form (183.8$\times$).}
\label{figure:case_crc}
\end{figure*}

\begin{figure*}[t]
\begin{promptbox}{Q\#72153725 $\cdot$ NumPy $\cdot$ 167.4$\times$ slowdown}
\textbf{Task.} Sum all $3{\times}3$ neighborhoods in a 2D array.\\[2pt]
\textbf{Input.} \texttt{'array'}: 2D int array $(\text{rows}, \text{cols})$ with rows, cols $\geq 3$.\\
\textbf{Output.} 2D float array $(\text{rows}{-}2, \text{cols}{-}2)$ of neighborhood sums.
\end{promptbox}

\begin{promptbox}{Slow (Qwen2.5-Coder-14B)}
\begin{Verbatim}[fontsize=\small, breaklines=true]
import numpy as np

def solution(input_data=None):
    array = input_data['array']
    rows, cols = array.shape
    result = np.zeros((rows-2, cols-2), dtype=array.dtype)
    for i in range(rows-2):
        for j in range(cols-2):
            result[i, j] = np.sum(array[i:i+3, j:j+3])
    return result
\end{Verbatim}
\end{promptbox}

\begin{promptbox}{Fast (Qwen3.5-397B)}
\begin{Verbatim}[fontsize=\small, breaklines=true]
import numpy as np

def solution(input_data=None):
    array = input_data['array']
    result = (array[:-2, :-2] + array[:-2, 1:-1] + array[:-2, 2:]
            + array[1:-1, :-2] + array[1:-1, 1:-1] + array[1:-1, 2:]
            + array[2:, :-2] + array[2:, 1:-1] + array[2:, 2:])
    return result.astype(float)
\end{Verbatim}
\end{promptbox}
\caption{Loop-to-Vectorization: per-cell \texttt{np.sum} loop vs.\ nine-slice vectorization (167.4$\times$).}
\label{figure:case_l2v}
\end{figure*}

\begin{figure*}[t]
\begin{promptbox}{Q\#68592930 $\cdot$ Pandas $\cdot$ 95.5$\times$ slowdown}
\textbf{Task.} Insert $n$ copies of a fixed row after every row in a DataFrame.\\[2pt]
\textbf{Input.} \texttt{'df'}: DataFrame $(R, 3)$; \texttt{'myrow'}: single-row DataFrame; \texttt{'n\_repeats'}: integer.\\
\textbf{Output.} DataFrame $(R \cdot (1{+}n), 3)$ interleaving originals and copies.
\end{promptbox}

\begin{promptbox}{Slow (GPT-5.4-mini)}
\begin{Verbatim}[fontsize=\small, breaklines=true]
import pandas as pd
import numpy as np

def solution(input_data=None):
    df = input_data["df"]
    myrow = input_data["myrow"]
    n_repeats = int(input_data["n_repeats"])
    n = len(df)
    labels = np.tile(np.arange(n_repeats + 1), n)
    df_repeated = df.loc[
        df.index.repeat(n_repeats + 1)
    ].reset_index(drop=True)
    myrow_block = pd.concat(
        [myrow] * (n * n_repeats), ignore_index=True)
    repeated_positions = labels != 0
    df_repeated.loc[
        repeated_positions, df.columns
    ] = myrow_block.to_numpy()
    return df_repeated
\end{Verbatim}
\end{promptbox}

\begin{promptbox}{Fast (DeepSeek-V4-Flash)}
\begin{Verbatim}[fontsize=\small, breaklines=true]
import numpy as np
import pandas as pd

def solution(input_data=None):
    df = input_data['df']
    myrow = input_data['myrow']
    n = input_data['n_repeats']
    N = len(df)
    total = N * (1 + n)
    cols = df.columns.tolist()
    data = {}
    for col in cols:
        data[col] = np.full(
            total, myrow[col].iloc[0],
            dtype=df[col].dtype)
    orig_indices = np.arange(N) * (1 + n)
    for col in cols:
        data[col][orig_indices] = df[col].values
    return pd.DataFrame(data)
\end{Verbatim}
\end{promptbox}
\caption{Data Layout Optimization: DataFrame replication vs.\ pre-allocated NumPy arrays (95.5$\times$).}
\label{figure:case_dlo}
\end{figure*}

\begin{figure*}[t]
\begin{promptbox}{Q\#69702937 $\cdot$ PyTorch $\cdot$ 218.4$\times$ slowdown}
\textbf{Task.} Compute the row-wise dot product of two square matrices.\\[2pt]
\textbf{Input.} \texttt{'A'}: torch.Tensor $(N, N)$; \texttt{'B'}: torch.Tensor $(N, N)$.\\
\textbf{Output.} torch.Tensor $(N,)$ where element $i = \sum_j A[i,j] \cdot B[i,j]$.
\end{promptbox}

\begin{promptbox}{Slow (GPT-5.4)}
\begin{Verbatim}[fontsize=\small, breaklines=true]
def solution(input_data=None):
    import torch
    if input_data is None:
        A = torch.tensor([[1, 2], [3, 4]])
        B = torch.tensor([[1, 1], [1, 1]])
    else:
        A = input_data["A"]
        B = input_data["B"]
        if not isinstance(A, torch.Tensor):
            A = torch.tensor(A)
        if not isinstance(B, torch.Tensor):
            B = torch.tensor(B)
    result = (A * B).sum(dim=1)
    return result
\end{Verbatim}
\end{promptbox}

\begin{promptbox}{Fast (Qwen3.5-397B)}
\begin{Verbatim}[fontsize=\small, breaklines=true]
import torch

def solution(input_data=None):
    A = input_data['A']
    B = input_data['B']
    if not isinstance(A, torch.Tensor):
        A = torch.tensor(A, dtype=torch.float32)
    if not isinstance(B, torch.Tensor):
        B = torch.tensor(B, dtype=torch.float32)
    result = (A * B).sum(dim=1)
    return result
\end{Verbatim}
\end{promptbox}
\caption{Overhead-Aware Stack Matching: function-level import and dead-code branches (218.4$\times$).}
\label{figure:case_osm}
\end{figure*}

\begin{figure*}[t]
\begin{promptbox}{EL w/ Rx System Prompt}
\begin{Verbatim}[fontsize=\scriptsize, breaklines=true]
## Role
You are an expert in data-science (DS) code performance optimization.
You will receive a correct but potentially slow Python solution along with its execution time profile.
Your task is to produce an optimized version that is functionally equivalent but faster.

---
## Optimization Guide:

Below is a reference optimization guide.
Use it as a checklist when analyzing the profile, but always verify that any change preserves correctness.

### 1. Loop-to-Vectorization (L2V)
When to apply: The profile shows hot Python for/while loops, iterrows(), apply(axis=1), or per-element indexing.
Key sub-patterns:
- Bulk Array Operation: Replace element-wise Python loops with NumPy/PyTorch vectorized ops.
- Table-Native Method: Replace iterrows/apply row loops with pandas/polars native methods.
- Shape Broadcasting: Replace nested scalar math with broadcasting or einsum.

### 2. Optimized Primitive Utilization (OPU)
When to apply: The code manually reimplements logic that a library already provides, or uses a slow API path.
Key sub-patterns:
- Direct Primitive Call: Replace hand-written logic with a single optimized library function.
- Lightweight API Path: Switch to a faster option within the same API family.
- Thin Dispatch Selection: Use the lighter backend variant (e.g., cKDTree over KDTree).

### 3. Data Layout Optimization (DLO)
When to apply: The code creates large unnecessary intermediate copies or uses a heavier container than needed.
Key sub-patterns:
- In-Place / View Reuse: Avoid allocating large temporaries; use views, inplace=True, or buffer reuse.
- Fit-for-Purpose Container: Match the container to the task (ndarray for numeric, dict for small lookups).
- Precise Dtype Selection: If suitable, use float32 over float64, Categorical for repeated strings.

### 4. Complexity-Reducing Computation (CRC)
When to apply: The code uses brute-force scans or generic solvers when a specialized algorithm exists.
Key sub-patterns:
- Pass Fusion / Early Exit: Merge multiple data passes into one, or terminate early.
- Index-Based Lookup: Replace linear scans with np.searchsorted, prefix sums, or binary search.
- Closed-Form Substitution: Replace iterative routines with algebraic identities.

### 5. Overhead-Aware Stack Matching (OSM)
When to apply: The code invokes a heavy framework for a task too small or simple to benefit from the overhead.
Key sub-patterns:
- Minimal-Path Execution: Remove unnecessary validation, defensive wrappers, or fallback logic.
- Native Python for Small Scale: Use dict/list/set instead of DataFrame/groupby when data is small.
- Interpreted Sufficiency: Skip Numba/JIT when NumPy vectorized ops already suffice.
---
Important: Always prioritize correctness over efficiency.
\end{Verbatim}
\end{promptbox}

\vspace{-2mm}

\begin{promptbox}{EL w/ Rx User Prompt}
\begin{Verbatim}[fontsize=\scriptsize, breaklines=true]
Here is a correct Python solution and its line-level execution time profile.
Your goal is to produce an optimized version that is functionally equivalent and significantly faster.

## Current Solution
```python
{code}
```

## Execution Time Profile
```
{profile}
```

## Your Task
Step 1 — Understand: Read the solution and its profile. Identify what the code does.
Step 2 — Diagnose: Locate the performance bottleneck(s) from the profile.
Step 3 — Plan: Based on your diagnosis, decide on an optimization strategy using the patterns above.
Step 4 — Optimize: Rewrite the solution. Ensure the same output, same signature: def solution(input_data=None).

Output your optimized Python code inside a single ```python``` code block.
\end{Verbatim}
\end{promptbox}
\caption{Full prompt for EL w/ Rx.}
\label{figure:prompt_el_rx}
\end{figure*}

\clearpage
\clearpage
\section{Evaluation Protocol Details}
\label{section:appendix_a3}

\subsection{Evaluation Protocol}
\label{appendix:evaluation_protocol}

\textbf{Execution Environment.}
All correctness and efficiency measurements are conducted on a dedicated server with the following specifications: dual Intel Xeon Platinum 8468V processors (96~physical cores, 192~threads, base 2.0\,GHz / turbo 3.8\,GHz), 512\,GB DDR5 RAM, and 4~NVIDIA H200 NVL GPUs (140\,GB each), running Ubuntu 22.04 LTS.

The evaluation environment uses Python~3.12 with the following library versions: NumPy~2.2, Pandas~2.3, PyTorch~2.10, SciPy~1.17, Polars~1.37, Numba~0.61, TensorFlow~2.20, and Scikit-learn~1.8. The server is used exclusively for evaluation with no concurrent workloads, ensuring stable execution time measurements.


\textbf{Test Case Generation.}
Each task contains on average 5.4 correctness sub-domains and 2.4 stress sub-domains. We set $N{=}5$ unique test cases per domain, yielding approximately 26.8 correctness cases and 12.0 stress cases per task (38.8 total).

\begin{figure}[t]
\centering
\includegraphics[width=0.9\columnwidth]{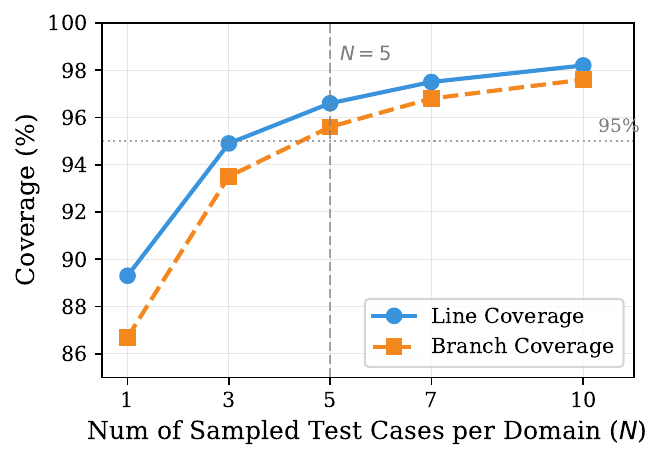}
\caption{Line and branch coverage of the reference solution. Both metrics exceed 95\% at $N{=}5$.}
\label{figure:seed_coverage}
\end{figure}

Concretely, the \texttt{input\_generator(seed, input\_domain)} takes a random seed and a domain identifier and returns a single test input. We iterate over all domains, starting from seed~42 and incrementing, collecting up to $N{=}5$ unique inputs per domain (duplicates are discarded). To validate this choice, we measure line and branch coverage of the reference solution as a function of $N$ on 100 randomly sampled tasks (Figure~\ref{figure:seed_coverage}). Coverage improves sharply from $N{=}1$ to $N{=}5$ (+7.3\% line, +8.9\% branch), then plateaus: $N{=}5$ to $N{=}7$ gains only +0.9\%/+1.2\%. At $N{=}5$, both metrics exceed 95\%, balancing coverage with evaluation cost across 16 models and 1{,}000 tasks.

\textbf{Execution Time Measurement Configuration.} Each test case is subject to a 5-second timeout; exceeding it is treated as a correctness failure. We adopt a repeated-execution scheme parameterized by $W$ (untimed warm-up runs) and $M$ (timed measurement runs, whose mean is reported). To select $W$ and $M$, we conduct a calibration experiment on 100 randomly sampled tasks: for each, we execute the reference solution under every $(W, M)$ combination with $W \in \{1, 3, 5\}$ and $M \in \{1, 2, 3, 4, 5\}$, and compute the coefficient of variation (CV) of the reported mean across 10 independent repetitions.

\begin{table}[t]
\centering
\scriptsize
\caption{Mean CV (\%) of reported mean execution time across $(W, M)$ configurations.}
\label{table:protocol_calibration}
\begin{tabular}{@{}lccccc@{}}
\toprule
& $M=1$ & $M=2$ & $M=3$ & $M=4$ & $M=5$ \\
\midrule
$W=1$ & 7.16\% & 4.92\% & 3.92\% & 3.34\% & 2.93\% \\
$W=3$ & 6.73\% & 4.62\% & 3.67\% & 3.12\% & 2.74\% \\
$W=5$ & 5.82\% & 3.98\% & 3.15\% & 2.67\% & 2.33\% \\
\bottomrule
\end{tabular}
\end{table}

Table~\ref{table:protocol_calibration} yields two observations that guide our choice. First, increasing $M$ yields the dominant variance reduction: from $M{=}1$ to $M{=}3$, CV drops by 3.24\% at $W{=}1$, whereas each subsequent increment gains less than 0.6\%. Second, increasing $W$ provides only marginal benefit: at fixed $M{=}3$, moving from $W{=}1$ to $W{=}5$ reduces CV by 0.77\% at the cost of four additional executions per test.

We select $W{=}1, M{=}3$ to balance measurement stability with cost. At our scale (16~models $\times$ 2~settings $\times$ 1{,}000~tasks $\times$ 38.8~cases), each additional execution per case adds ${\sim}$1.2M runs. A CV of 3.92\% is well within the discriminative range.

\textbf{Domain-Level Execution Time Statistics.} We summarize the execution time distribution of the reference solution across the two domain groups. On correctness domains, the reference solution runs in 1.80\,ms at the median, 20.84\,ms on average, and 40.24\,ms at the 90th percentile (P90). On stress domains it runs substantially longer, at 25.12\,ms median, 132.56\,ms mean, and 199.44\,ms P90. The median stress-domain execution time of 25.12\,ms provides sufficient headroom for discriminating efficient from inefficient candidate solutions.

\subsection{Evaluated Models}
\label{appendix:evaluated_models}

We evaluate 16~models grouped into three accessibility tiers. Closed Frontier models are accessed via proprietary APIs; Open Frontier and Open Coder models are open-weight and served via vLLM. Below we list each model by its display name followed by the official model identifier used in our experiments.

The \textbf{Closed Frontier} tier includes two models from OpenAI~\citep{openai2025gpt54}, GPT-5.4 (gpt-5.4) and GPT-5.4-mini (gpt-5.4-mini), and two from Anthropic~\citep{anthropic2025claude4}, Claude-Opus-4.6 (claude-opus-4-6) and Claude-Sonnet-4.6 (claude-sonnet-4-6).

The \textbf{Open Frontier} tier spans five general-purpose models: DeepSeek-V4-Flash (deepseek-ai/DeepSeek-V4-Flash)~\citep{deepseek2026v4flash} and DeepSeek-V3.2 (deepseek-ai/DeepSeek-V3.2)~\citep{liu2024deepseek} from DeepSeek, Kimi-K2.5 (moonshotai/Kimi-K2.5)~\citep{team2026kimi} from Moonshot AI, MiniMax-M2.5 (MiniMaxAI/MiniMax-M2.5)~\citep{minimax2025m25} from MiniMax, and Qwen3.5-397B/A17B (Qwen/Qwen3.5-397B-A17B)~\citep{yang2025qwen3} from Alibaba Qwen.

The \textbf{Open Coder} tier covers seven code-specialized models: Qwen3-Coder-Next (Qwen/Qwen3-Coder-Next)~\citep{cao2026qwen3}, Qwen2.5-Coder-32B (Qwen/Qwen2.5-Coder-32B-Instruct), and Qwen2.5-Coder-14B (Qwen/Qwen2.5-Coder-14B-Instruct)~\citep{hui2024qwen2} from Alibaba Qwen; Devstral-2-123B (mistralai/Devstral-2-123B-Instruct-2512)~\citep{mistral2025devstral2} and Codestral-22B (mistralai/Codestral-22B-v0.1)~\citep{mistral2024codestral} from Mistral AI; and DeepSeek-Coder-33B (deepseek-ai/deepseek-coder-33b-instruct)~\citep{guo2024deepseek} and DeepSeek-Coder-V2-Lite (deepseek-ai/DeepSeek-Coder-V2-Lite-Instruct)~\citep{zhu2024deepseek} from DeepSeek.

\end{document}